\documentclass[aps,twocolumn,secnumarabic,groupedaddress,floatfix,amsfonts]{revtex4}
\usepackage[dvips]{graphicx,color}

\begin{document}
\title{Collider Inclusive Jet Data and the Gluon Distribution}
\author { Jon Pumplin$^a$, J.~Huston$^a$, H.~L.~Lai$^b$,
P.~M.~Nadolsky$^d$, Wu-Ki Tung$^{a,c,e}$ and C.-P.~Yuan$^a$}
\affiliation{
$^a$Michigan State University, East Lansing, Michigan \\
$^b$Taipei Municipal University of Education, Taipei, Taiwan \\
$^c$University of Washington, Seattle, Washington;
$^e$Deceased \\
$^d$Department of Physics, Southern Methodist University, Dallas, Texas}
\date{\today}
\preprint{MSUHEP-090415}
\begin{abstract} 
Inclusive jet production data are important for constraining the gluon 
distribution in the global QCD analysis of parton distribution functions.  
With the addition of recent CDF and D0 Run II jet data, we study a number 
of issues that play a role in determining the up-to-date gluon distribution 
and its uncertainty, and produce a new set of parton distributions that 
make use of that data. We present in detail the general procedures used 
to study the compatibility between new data sets and the previous body 
of data used in a global fit.  We introduce a new method in which the 
Hessian matrix for uncertainties is ``rediagonalized'' to obtain 
eigenvector sets that conveniently characterize the uncertainty of a 
particular observable.  
\end{abstract}
\pacs{12.38.Qk, 12.38.Bx, 13.60.Hb, 13.85.Qk}
\maketitle

\section{Introduction\label{sec:intro}}

The gluon distribution $g(x,\mu)$ plays an important role in high
energy collider phenomenology, both for standard model and new 
physics. Yet it is the most elusive of the parton distribution 
functions (PDFs) in contemporary global QCD analysis.  
At moderate values of the momentum fraction $x$, extensive high
precision data on deep inelastic scattering (DIS) constrain 
$g(x,\mu)$ fairly well through the $\mu$-dependence 
that is predicted by QCD. 
The little information we have at large $x$ comes mostly from inclusive 
jet production at hadron colliders, which receives contributions 
directly from the gluon distribution at leading order in $\alpha_s$.
The recently published inclusive jet data from 
Tevatron Run II measurements by CDF \cite{cdfR2} and D0 \cite{D0R2} 
are therefore of considerable interest for improving our knowledge 
of the gluon distribution.  

Previous CTEQ studies \cite{cteq606165,cteq66} have used only the 
Run I jet data \cite{cdfR1,D0R1}.  A recent MSTW study \cite{MSTW2008} 
includes the Run II data in an analysis with aims parallel to this one.  
A comparison with their results is presented in Sec.\ \ref{sec:CompareMSTW}.

In this paper, we make a detailed study 
of several issues that bear on the behavior of the gluon distribution and 
its range of uncertainties, focusing on the use of Tevatron inclusive jet 
data.  (Inclusive jet production in DIS processes can also 
provide constraints on the gluon distribution; but those constraints are
considerably weaker and we do not include them here.) 
Some of the results and techniques described here are known to many 
practitioners in the field, but have not been previously documented in 
the literature.  Some of these results are frequently misunderstood---e.g., 
in discussions at workshops---so it seems worthwhile to set them out in 
systematic detail.  
The methods discussed here for the inclusive jet data thus serve as a 
pedagogical study of techniques that can be applied in general when new 
data sets become available to advance the PDF analysis.  

One of the techniques we use is presented here for the first time.  It 
involves orienting the choice of eigenvector directions in the Hessian 
method in order to simplify the study of uncertainty for any
particular quantity of interest.

\begin{figure*}[tbh]
\begin{center}
\mbox{
 \resizebox*{0.42\textwidth}{!}{
\includegraphics[clip=true,scale=1.0]{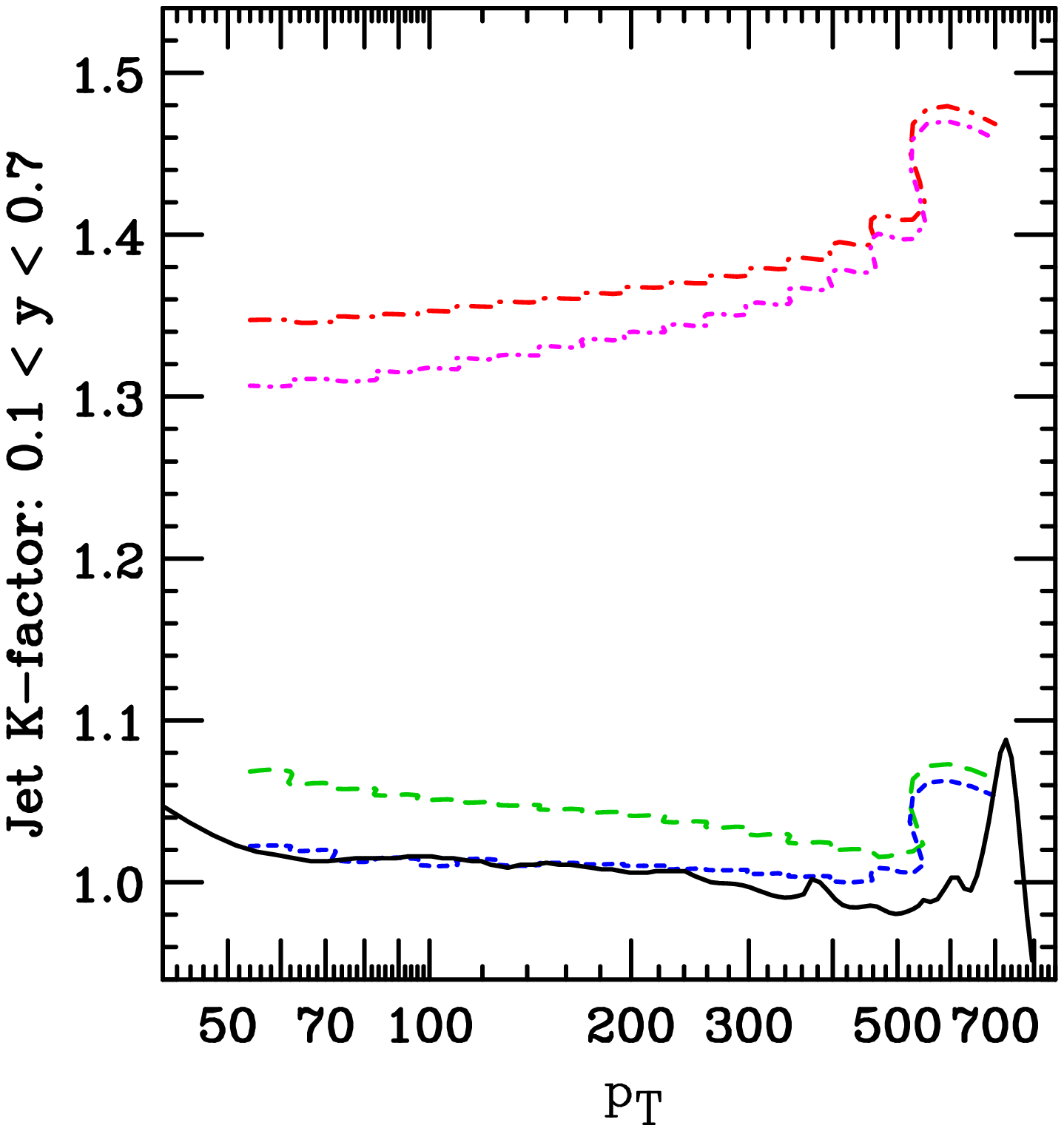}}
\hfill
 \resizebox{0.42\textwidth}{!}{
\includegraphics[clip=true,scale=1.0]{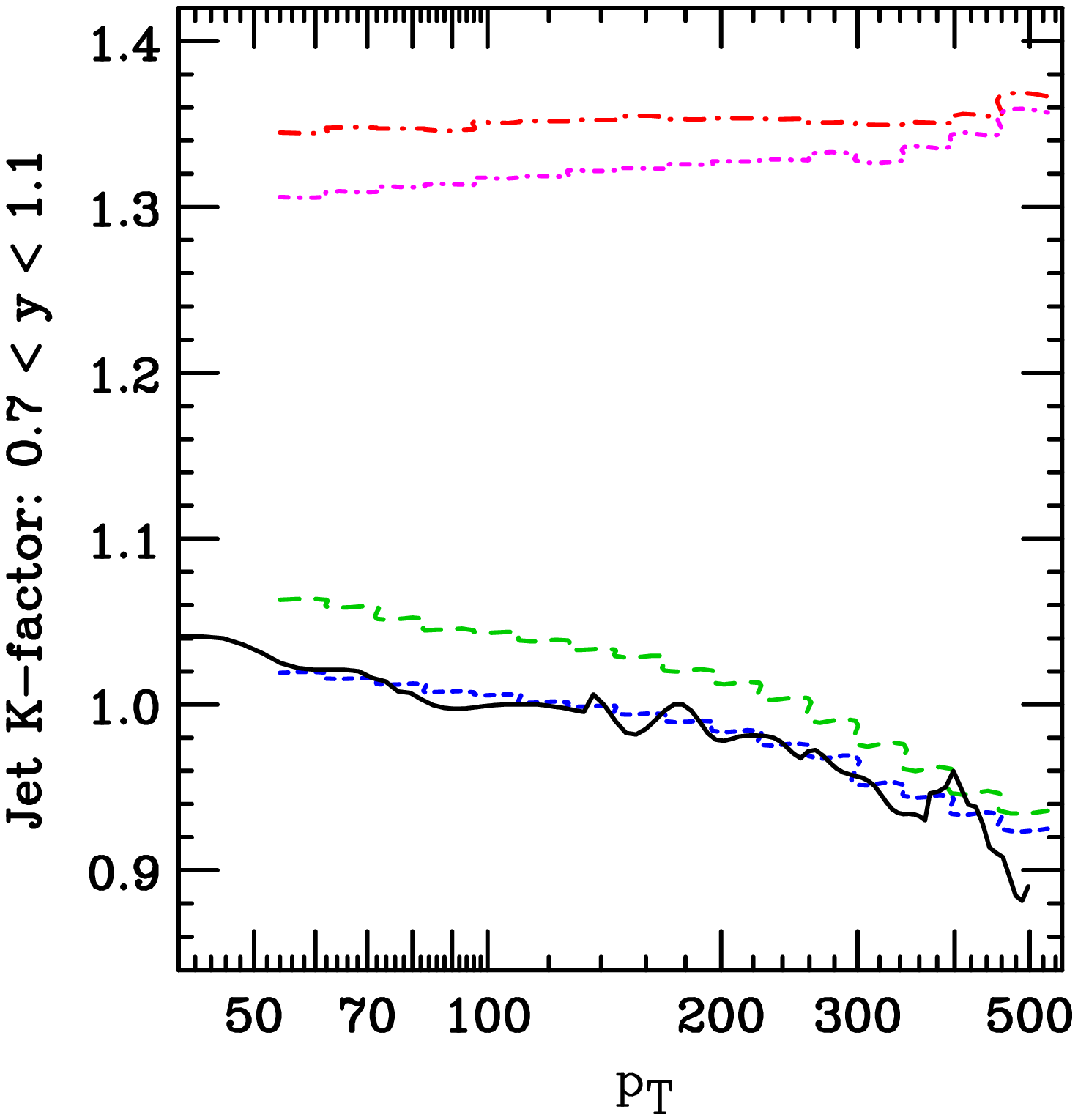}}
}
\end{center}
\begin{center}
\mbox{
\resizebox{0.42\textwidth}{!}{
\includegraphics[clip=true,scale=1.0]{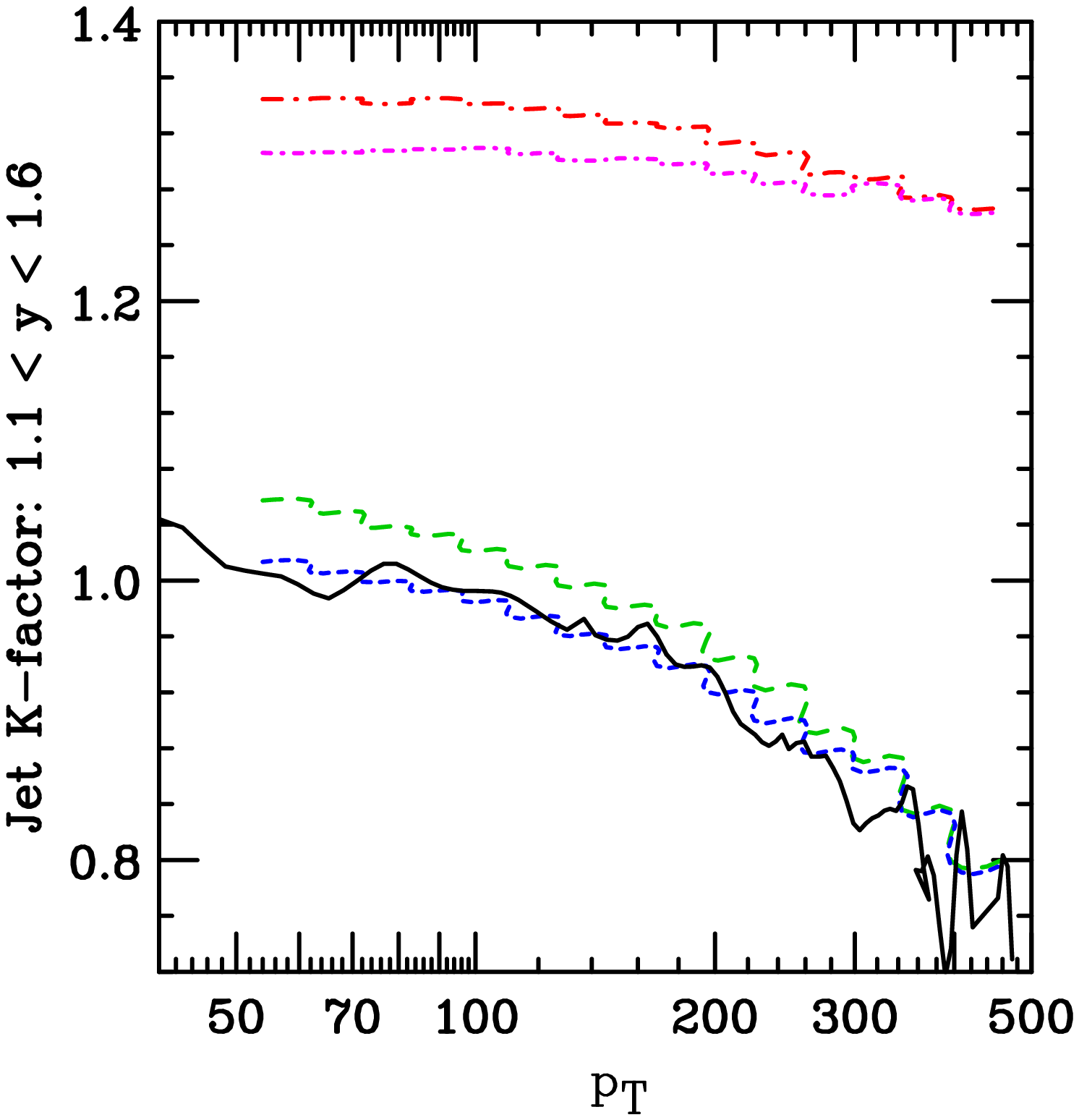}}
\hfill
\resizebox{0.42\textwidth}{!}{
\includegraphics[clip=true,scale=1.0]{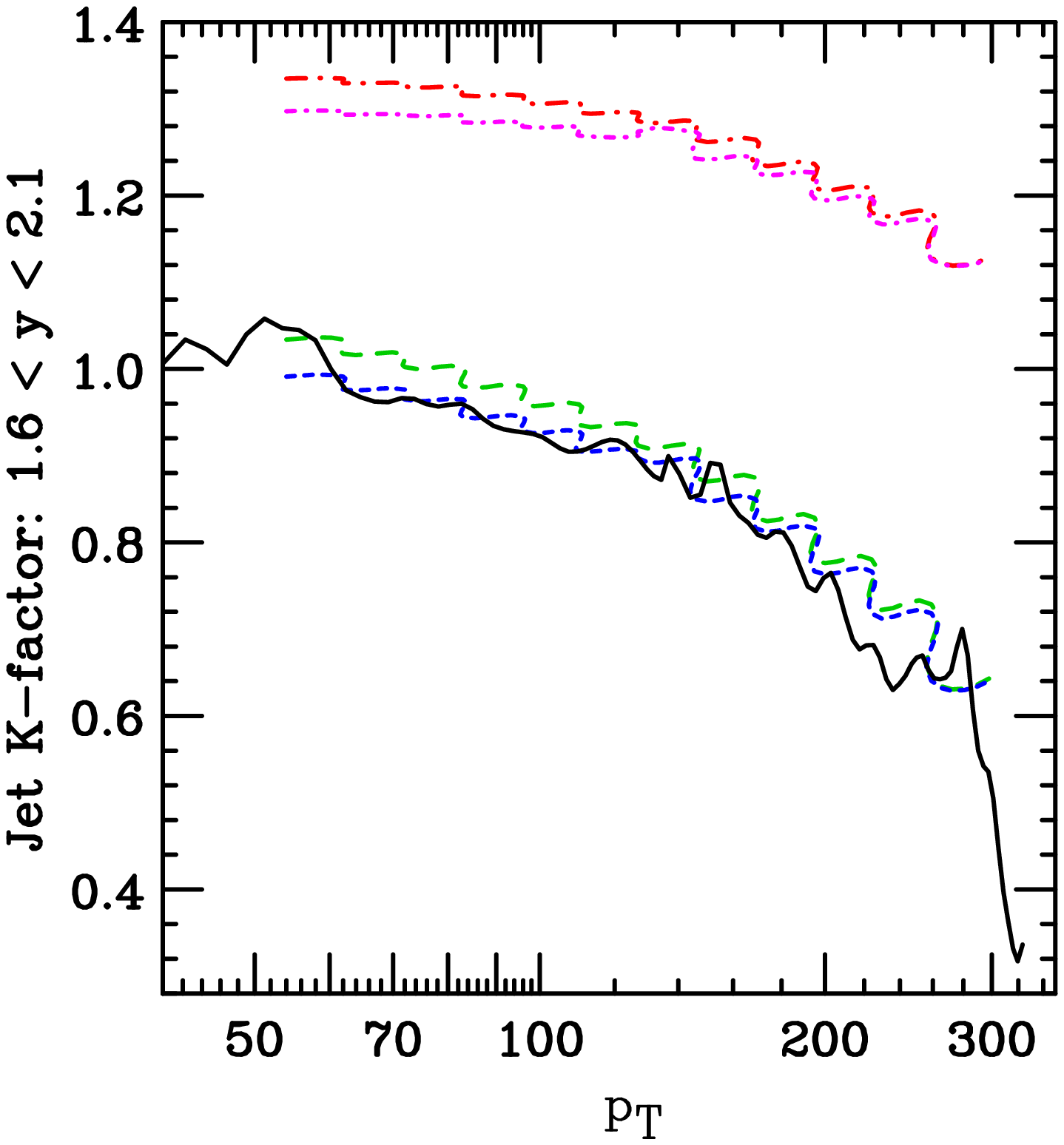}}
}
\end{center}
 \caption{Theory calculations for the ratio K = NLO/LO from 
FastNLO and EKS. 
FastNLO with $\mu \! = \! p_T$:   
$R_{\mathrm{sep}}\! =\! 2.0$ (long dash dot), 
$R_{\mathrm{sep}}\! =\! 1.3$ (short dash dot);
FastNLO with $\mu \! =\!  p_T/2$: 
$R_{\mathrm{sep}}\! =\! 2.0$ (long dash), 
$R_{\mathrm{sep}}\! =\! 1.3$ (short dash);
EKS with $\mu \! =\!  p_T/2$, $R_{\mathrm{sep}}\! =\! 1.3$ (solid).
}
 \label{fig:figOne}
\end{figure*}

\section{Theory calculations for inclusive jets\label{sec:theory}}

Up to now, the CTEQ global analyses of jet cross sections as a function 
of jet transverse momentum $p_T$ have been
based on the EKS NLO program \cite{EKS}. Recently, the FastNLO
implementation \cite{FastNlo} of the NLOJET++ \cite{NloJet}
calculation has gained increasing use---in part because of its convenient
interface.
(FastNLO allows the dependence on the PDFs of the NLO cross section 
to be included in the computation of $\chi^2$ at every 
step within the fitting procedure.  However, we find that calculating 
the ratio K=NLO/LO for each data point using a single typical fit to the 
data provides an adequate approximation.)
To make sure that the two calculations are consistent in the
global analysis context, we have directly compared their results 
in the Tevatron Run I and II kinematic ranges. The
theoretical results also depend on choices of: (i) the renormalization
and factorization scales in PQCD, usually taken to be the same, say
$\mu$; and (ii) the jet algorithm, including parameters such as 
$R_{\mathrm{sep}}$ (for separation of neighboring jets) \cite{cdfR2}. 
We have performed 
the comparison using a variety of these choices. The results provide 
information on the importance of these factors for the global analysis.

\begin{figure*}[tbh]
\begin{center}
\mbox{
 \resizebox*{0.42\textwidth}{!}{
\includegraphics[clip=true,scale=1.0]{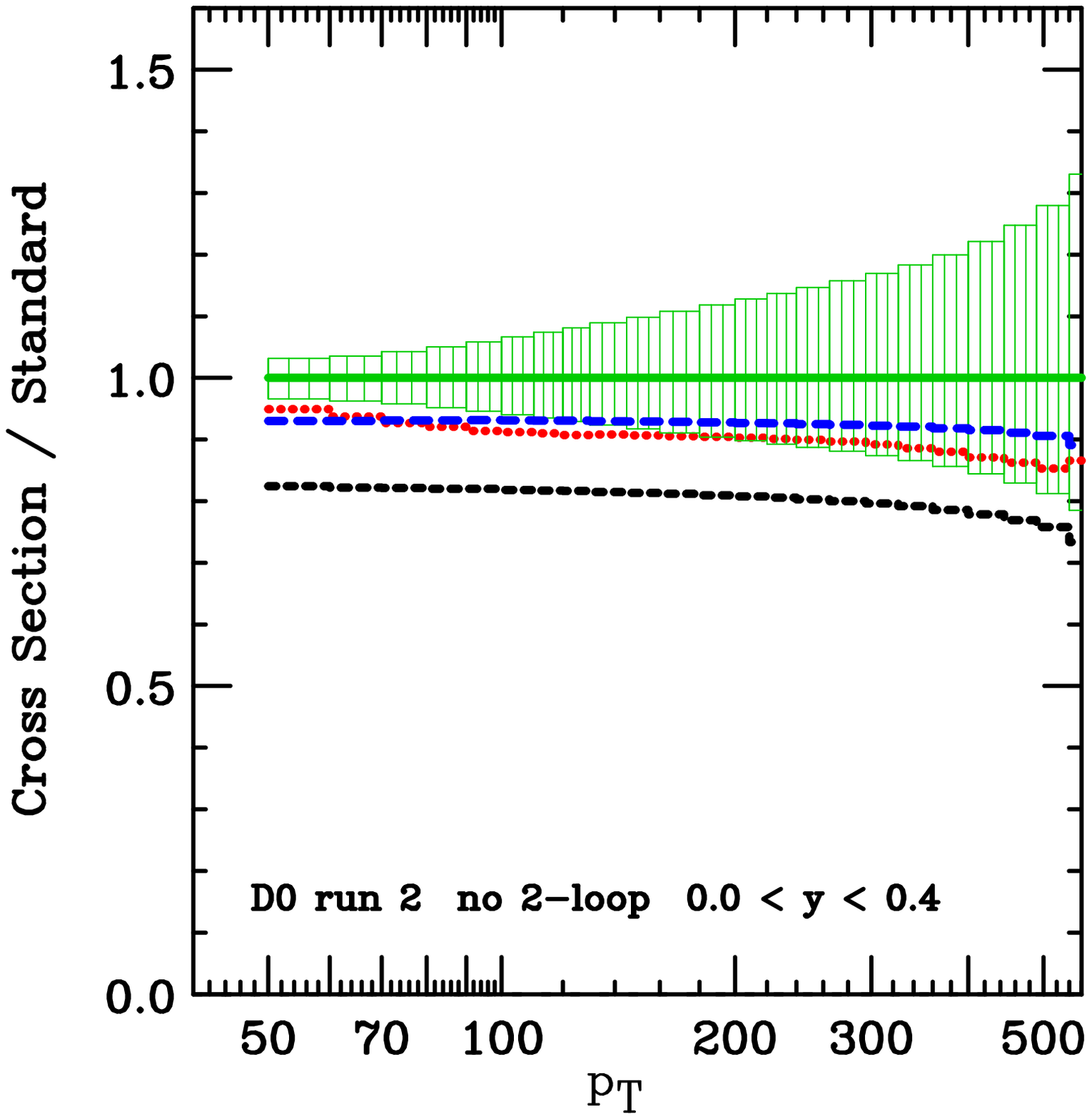}}
\hfill
 \resizebox{0.42\textwidth}{!}{
\includegraphics[clip=true,scale=1.0]{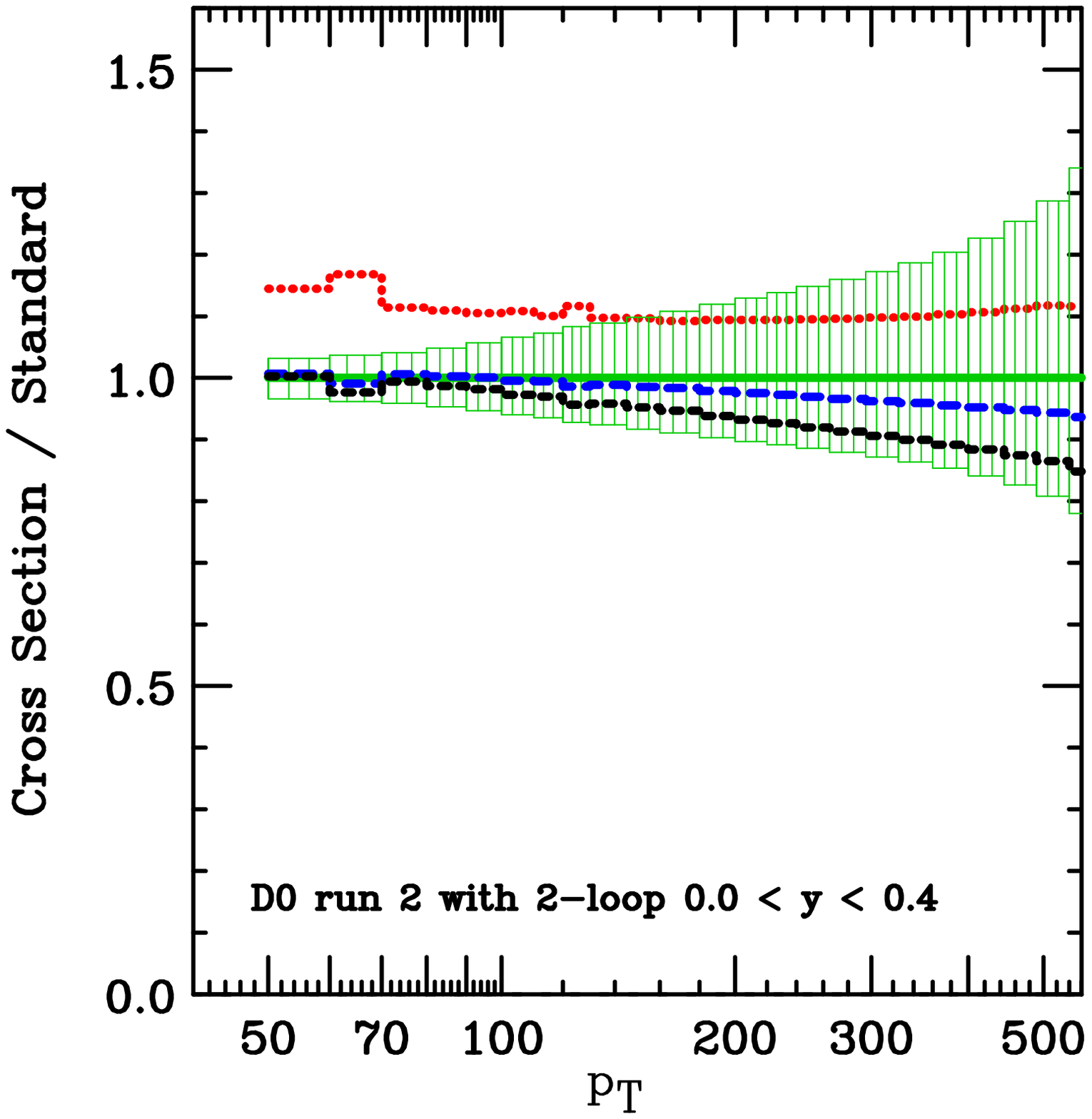}}
}
\end{center}
\begin{center}
\mbox{
\resizebox{0.42\textwidth}{!}{
\includegraphics[clip=true,scale=1.0]{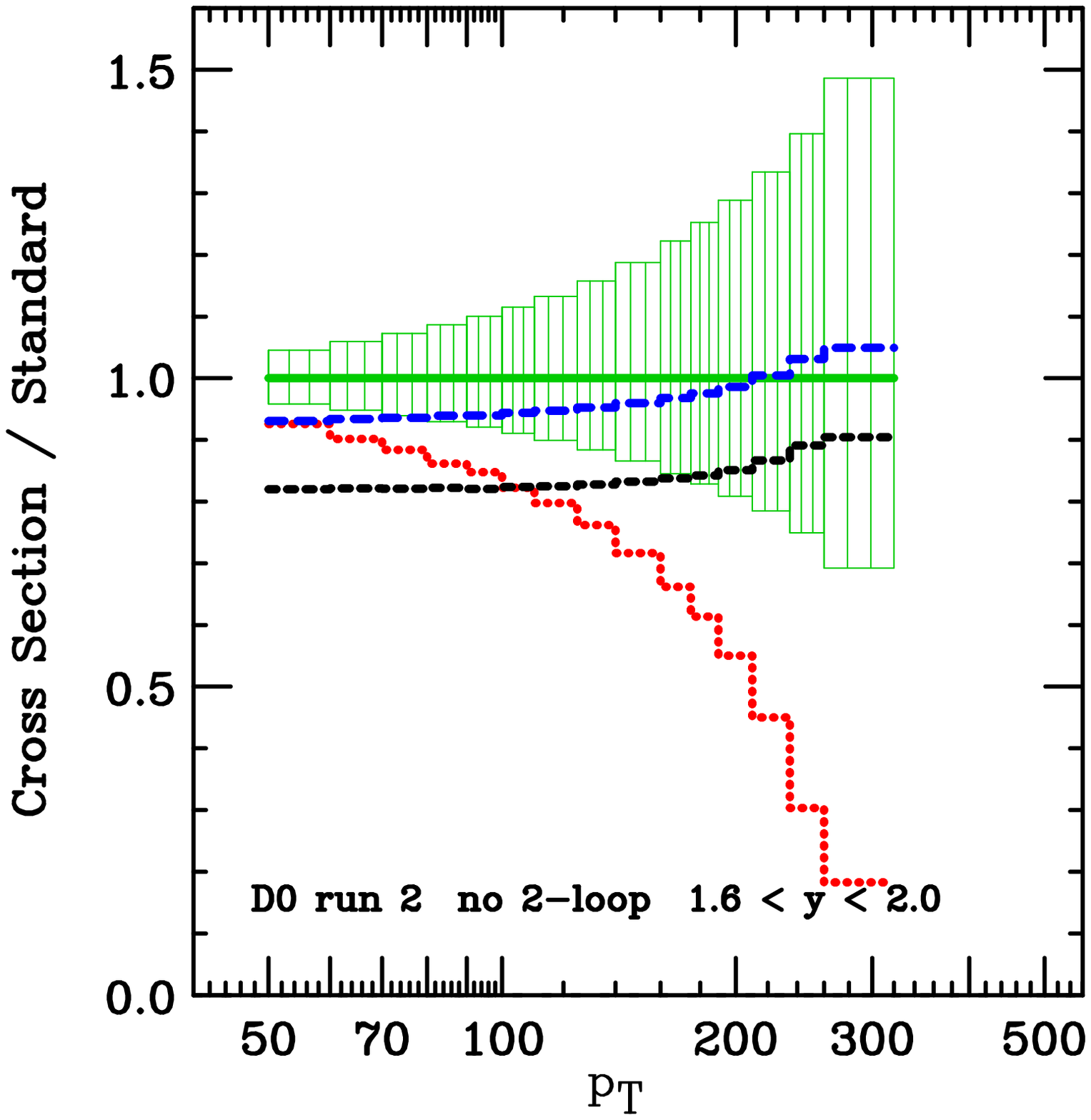}}
\hfill
\resizebox{0.42\textwidth}{!}{
\includegraphics[clip=true,scale=1.0]{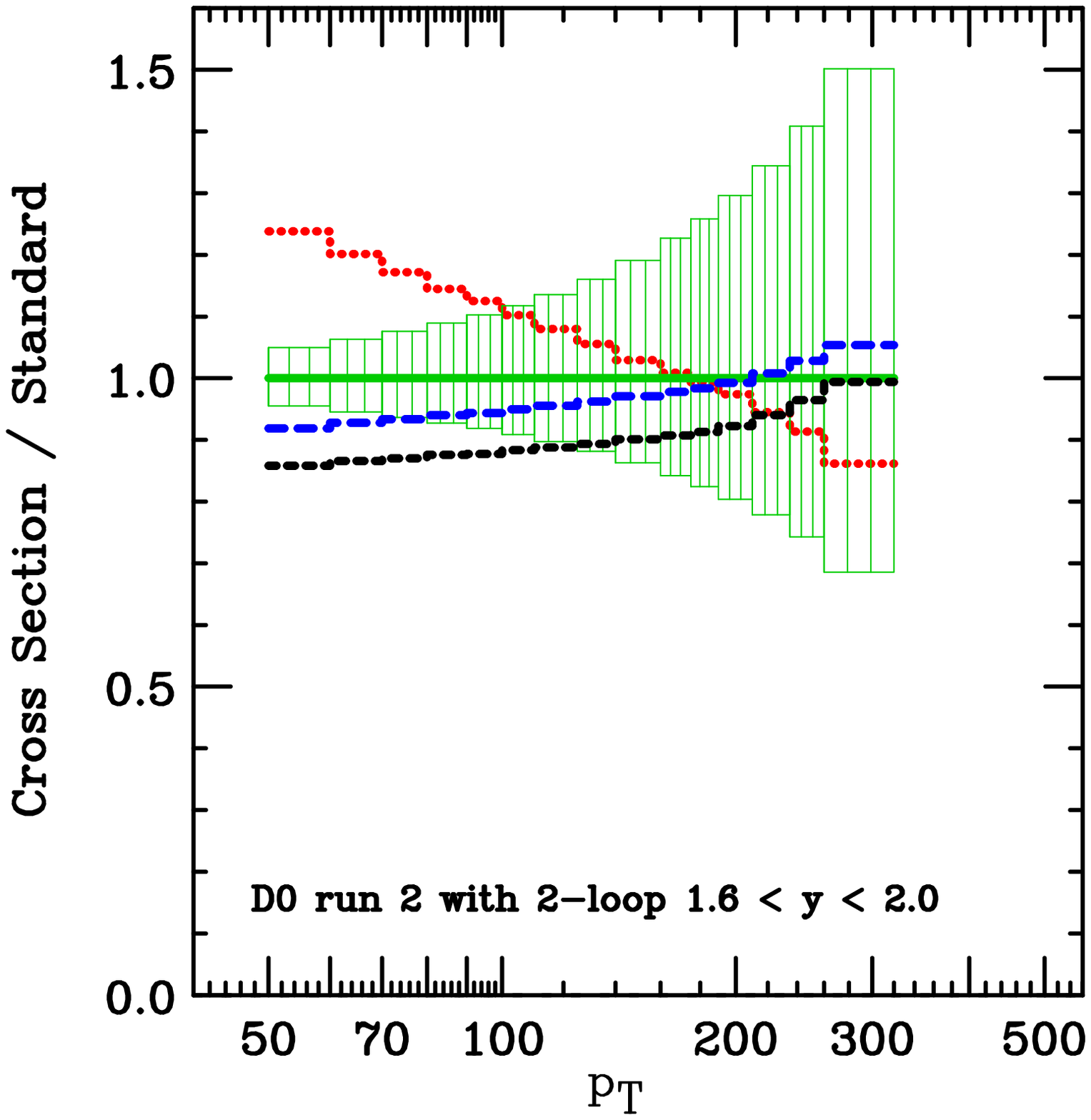}}
}
\end{center}
 \caption{Effect of scale choice on predicted cross section with 
$R_{\mathrm{sep}}=1.3$:
$\mu = 2\,p_T$ (short dash), $p_T$ (long dash), $p_T/2$ (solid), $p_T/4$ (dotted),
relative to our Standard Choice ($\mu = p_T/2$, $R_{\mathrm{sep}}=1.3$,
no ``two-loop'' correction).
Right panels include the ``two-loop'' resummation correction.
Uncertainty bands from PDFs are shown for comparison.
}
 \label{fig:figTwo}
\end{figure*}

Figure \ref{fig:figOne} shows the K-factor, defined by 
K = NLO/LO, for the Tevatron Run II $p_{T}$
range in several of the experimental rapidity intervals.  
Each plot shows results from FastNLO for two choices of the scale:
$\mu =  p_T$ (upper two curves), and 
$\mu = p_{T}/2$ (lower two dashed curves).  
Within each of these pairs, 
the upper curve uses the midpoint cone jet algorithm with 
$R_{\mathrm{sep}}=2.0$, while the lower curve 
uses the midpoint algorithm with $R_{\mathrm{sep}}=1.3$.  The solid 
curve shows the result of the EKS program for 
$\mu = p_{T}/2$ and $R_{\mathrm{sep}}=1.3$.
(The wiggles in this curve are caused by fluctuations from the 
Monte Carlo integration used in EKS.) We observe the 
following:

\noindent$\bullet$ 
The overall agreement between the EKS and FastNLO calculations is 
satisfactory, though not perfect. Our results from parallel 
global analyses based on these two methods for calculating jets, with 
all other options identical, show good agreement, which indicates 
that results of the global analysis are not sensitive to deviations 
of the magnitude shown. We use the FastNLO results
for the remainder of this investigation.

\noindent$\bullet$ 
The effect of $R_{\mathrm{sep}}$ choice is quite small.

Since the scale choice affects the predicted cross section directly through 
the LO cross section, as well as through the K factor, we explore it further 
in Fig.\ \ref{fig:figTwo}, which shows the predicted cross section for various 
scale choices normalized by our ``standard'' choice.
The plots on the left correspond to the conventional NLO calculation, while 
those on the right also include a ``2-loop'' correction derived from threshold 
resummation \cite{twoloop}, which is available in FastNLO.  
We only show results from one central and one 
large rapidity bin; results at intermediate $y$ interpolate 
between these two.
The bands in each plot represent the estimated uncertainty due to the 
PDFs, for comparison. 
We conclude the following:

\noindent $\bullet$ 
The low-scale choice $\mu =p_{T}/4$ leads to results that are far from 
the other choices at large rapidity, and shows unstable behavior with 
respect to 2-loop 
corrections, which lie mostly outside the PDF uncertainty bands. 
This scale choice is thus unsuitable for theoretical calculations, 
as is also apparent from the fact that K = NLO/LO is far from 
$1$ with this choice, which suggests that still higher order 
corrections would be very important.  By contrast, the other three scale 
choices show consistent patterns and yield stable results.

\noindent $\bullet$ 
One may use the range $p_{T}/2 < \mu < 2 \, p_{T}$ to empirically 
estimate the uncertainty due to uncalculated higher-order corrections. 
This range of theoretical uncertainty is seen to be 
fairly independent of $p_{T}$, in contrast to the uncertainty due to 
PDFs, which has a strong $p_{T}$ dependence. The theory uncertainty is 
comparable to the PDF uncertainty in the low $p_{T}$ range, but is much 
smaller than it in the high $p_{T}$ range.

\noindent $\bullet$ The theoretical uncertainties are reduced 
in the calculation that includes the partial 2-loop correction. Whether 
this reduction provides a genuine increase in accuracy depends on the 
reliability of the approximation, which is still controversial.  
We do not use this correction in the remainder of the paper.

With these theoretical background studies completed, we now proceed 
to study the impact of the Tevatron jet data on determining the gluon 
distribution.

\section{Preliminary global fit with the new jet data
\label{sec:fits}}

We use the published CTEQ6.6 PDF set \cite{cteq66} 
as the reference fit for our comparison study. 
Unless otherwise stated, the theoretical and experimental 
inputs are kept the same as in \cite{cteq66} except for the addition 
of the CDF \cite{cdfR2} and D0 \cite{D0R2} Run II data sets.  

We use the CDF Run II results obtained from the midpoint cone jet 
algorithm, rather than the earlier results based on the 
$k_T$ algorithm \cite{CDFKT}. The two analyses were carried
out on the same events, so it be incorrect to include them 
both; and the ratio of the resulting cross sections agrees well with 
the ratio  predicted by NLO QCD, as stated in \cite{cdfR2}.  
(The CDF data were supplied to us by one of the CDF authors, so we were 
not affected by errors in the original publication \cite{cdfR2}, which 
have now been corrected as described in its first reference.)

The CTEQ6.6 central fit and its eigenvector sets which characterize the 
uncertainty are known to describe the Run II jet data fairly well, 
even though those data were not available at the time of the CTEQ6.6 analysis. 
Thus from the outset we know that no revolutionary changes will result from 
incorporating the new data into the global analysis.
The purpose of our study is to quantify what 
changes there are; and to investigate some subtle features that have not
been explored before, which have implications for our efforts to
pin down the gluon PDF.

With the addition of the Tevatron Run II jet data, our global analysis
includes 37 data sets with a total of 2898 data points. 
As a baseline, when CTEQ6.6 PDFs are used directly to compute the cross
sections and then compared to these data points, we obtain a good
overall fit with $\chi^{2}=2756$.  
\emph{Here and in all our fits, the full correlated 
experimental errors are used in computing $\chi^{2}$, 
for all experiments that report their errors in this form.}
To get a first look at the impact of the Run II jet data, we performed 
a preliminary fit using the same theoretical input as CTEQ6.6. In 
this fit, the weighted $\chi^{2}$ becomes $2740$---a reduction of $16$.
This is a very small reduction when spread over all 2898 data 
points, or even when spread over the 182 new ones---as was anticipated 
since CTEQ6.6 already provided a reasonably good fit to the new data.  

The only significant change in the best-fit 
PDFs from CTEQ6.6 to the preliminary fit occurs in the gluon PDF. 
This can be demonstrated by repeating the global fit with 
all of the quark distribution parameters frozen at their CTEQ6.6 values, 
thus only allowing the gluon distribution to change.  The reduction 
in $\chi^{2}$ is nearly the same and the resulting fit is essentially
equivalent to the preliminary fit.  This confirms our expectation that the 
inclusive jet data provide a handle on $g(x,\mu)$ and little else.

\section{Parametrizing the gluon distribution}
\label{sec:parametrizing}

Since the jet data are sensitive to the gluon distribution, it is essential 
to use a sufficiently flexible parametrization for the gluon at the starting 
scale $\mu_0$ for DGLAP evolution, which we choose to be 
$1.3 \, \mathrm{GeV}$ as in previous analyses.  The form we use is 
\begin{equation}
g(x,\mu_0) = \\
     a_0 \, x^{a_1} \, (1 - x)^{a_2} \, 
     \exp(a_3 x \, + \, a_4 x^2 \, + \, a_5\sqrt{x}) \; .
  \label{eq:GluonParametrization}
\end{equation}
We add a penalty to the overall $\chi^2$ to force parameter $a_2$, 
which controls the behavior at $x \to 1$, to lie within a reasonable but 
generous range $0.5 \lesssim a_2 \lesssim 10$.  
The form (\ref{eq:GluonParametrization}) is more general than what was used in 
CTEQ6.6, which was equivalent to $a_2=4$ and $a_5=0$.
Alternative parametrizations have also been tested, to assure that our 
results are not sensitive to the particular form of smooth function 
that we choose to multiply the basic $ x^{a_1}$ and $(1 - x)^{a_2}$ factors.

Because the gluon distribution is not strongly constrained by existing data, 
it has been common to use a fairly restricted functional form for the 
nonperturbative input function, compared to the better-constrained light 
quark distributions. A frequent practice is to start with the minimal form
$x^{a_1}\,(1-x)^{a_2}$ and incrementally add new parameters until the 
quality of the global fit ceases to improve. This is 
a sensible approach for finding a reasonable ``best fit'' PDF set. 
But it can produce misleading results by artificially reducing the 
estimated uncertainties---as happened famously when the CDF Run I 
measurements of the jet cross section at first appeared to lie outside 
the range of standard model predictions at large $p_{T}$ \cite{cdfHJ,cteqHJ}.  
We will discuss related examples of this in 
Secs.\ \ref{subsec:RestrictedGluon} and \ref{sec:GluonUncertaintyResults}.

In current practice, the number of parameters used by
various groups shows wide variation, which depends both on the
constraining power of the input experiments included in the analysis
and on the stability of the analysis method.
Using too few parameters can lead to uncertainty estimates that 
reflect the assumed functional forms more than the experimental 
constraints.  This fact appears to be under-appreciated, since the 
number of parameters used for uncertainty studies is commonly kept 
at a minimum level based only on the central fit.

A technical reason to restrict the number of parameters in the uncertainty 
study is the instability of the Hessian method for determining the extreme 
PDF eigenvector sets, which occurs as
the number of fitting parameters approaches the 
limit of constraining power of the experimental input. 
We have been able to overcome this problem by using the iterative method 
developed in \cite{multivariate} to
control both the instabilities due to vast disparities in the 
unscaled eigenvalues and instabilities caused by the numerical 
evaluation of the second derivatives that define the Hessian matrix.
The set of tools we have developed provide an orderly way to obtain 
stable results as the number of parameters is increased. 
This is the reason why the CTEQ analyses have consistently used a larger 
number of uncertainty eigenvector sets than other groups.  The fits 
described in Sec.~\ref{sec:GluonUncertaintyResults} use 24 parameters 
to describe the PDFs at $\mu_0$.

A Neural Network approach to the PDF analysis (NNPDF) \cite{NNPDF} has been 
developed recently to circumvent the parametrization issue.  This appears 
highly promising.  However, to make this approach as effective as 
possible, it may be important to retain some theory-based guidelines 
on the PDFs at scale $\mu_0$.  In particular, there are good physical 
arguments behind the traditionally assumed behaviors $x^{a_1}$ at 
$x \to 0$ and $(1-x)^{a_2}$ at $x \to 1$, which even predict estimates 
for the constants $a_1$ and $a_2$ that one may wish to harness.  
The validity of those arguments is supported by the observation that for 
the $u$ quark distribution, which is the most accurately measured of the 
PDFs, the fitted results for $a_1$ and $a_2$ lie close to their 
theoretical expectations.

We now proceed to a detailed study of the compatibility of the jet data 
sets with each other and with the nonjet data.  This study also serves 
as a case study of methods to apply when adding new data sets to
a global fit.

\section{Testing compatibility of data sets using weighted $\mathbf{\chi^2}$}
\label{sec:TestingCompatibility}

When one contemplates adding new sets of experimental data to an existing 
global analysis, one begins by asking a series of questions that 
can be answered systematically by making fits in which the 
$\chi^2$ for the new data sets are multiplied by various weight factors.
These weight factors multiply the contributions from individual data sets
before they are added to the global $\chi^2$ that is minimized in the fit, 
in order to vary how much influence each set is allocated in determining the 
fit.  For a related discussion of these ideas, see \cite{CollinsPumplin}.

\medskip\noindent
\textit{Are the new data consistent with theory?} 
can be addressed in a minimal way by seeing if $\chi^2$ for the new data is 
acceptably close to its nominal range of 
$N \pm \sqrt{2\,N}$ for $N$ data 
points, at least when these data are assigned a sufficiently large weight.
(In the ideal situation of Gaussian experimental errors, 
this range corresponds to a $1\sigma$ confidence interval 
around the best-fit $\chi^2$. In the present case, 
where the bulk of the experimental error may come from
systematic effects, this comparison may also reveal deviations 
from Gaussian behavior, which are known to occur
when the experimental errors are predominantly systematic.)

\medskip\noindent
\textit{Are the new data consistent with the previous experiments?} can be 
addressed by observing the increase in $\chi^2$ for the original 
data that occurs when the fit is adjusted to accommodate the new data.

\medskip\noindent
\textit{Are the new data sets consistent with each other?}
can be studied by observing the change in $\chi^2$ for each new data set 
in response to changing the weights for the other new data sets.  This will 
reveal whether two new data sets ``pull'' in the same direction, or whether
on the contrary there is a ``tension'' between them; or whether they measure 
different features, and so have little effect on each other.

\medskip\noindent
\textit{Do the new data sets provide significant new constraints?} can be 
studied in a simple way by exploring the range of acceptable fits to the 
original data using the Hessian (eigenvector) method, and observing how 
many of these eigenvector sets produce acceptable fits to the new data.

\begingroup
\begin{table}[!hbt]
\squeezetable
  \begin{center}
\begin{tabular}{||rr|rr|rr|rr||r||}
\hline
\multicolumn{2}{||c|}{CDF$_{\mathrm{I}}$  N=33} & 
\multicolumn{2}{|c|}{D0$_{\mathrm{I}}$ N=90} &
\multicolumn{2}{||c|}{CDF$_{\mathrm{II}}$ N=72} & 
\multicolumn{2}{|c||}{D0$_{\mathrm{II}}$ N=110} &
$\Delta \chi^2$ \\ 
Wt & $\chi^{2}$ & Wt & $\chi^{2}$ & Wt & $\chi^{2}$ & Wt & $\chi^{2}$ 
& nonjet\\ 
\hline
 0 & 55.4 &  0 & 115.3 &  0 &  99.5 &  0 & 134.0 &   0.0 \\         
 1 & 52.6 &  1 &  47.0 &  0 & 105.6 &  0 & 138.3 &  11.8 \\         
 0 & 56.6 &  0 &  82.2 &  1 &  85.6 &  1 & 124.1 &   6.2 \\         
 1 & 52.1 &  1 &  59.4 &  1 &  88.5 &  1 & 121.5 &   9.6 \\ \hline  
 0 & 58.4 &  0 &  60.9 & 10 &  79.6 & 10 & 120.4 &  39.9 \\         
 1 & 54.8 &  1 &  58.8 & 10 &  80.3 & 10 & 120.0 &  39.4 \\         
10 & 54.1 & 10 &  35.6 &  0 & 112.9 &  0 & 156.7 &  24.1 \\         
10 & 53.1 & 10 &  38.6 &  1 & 102.6 &  1 & 142.3 &  21.9 \\         
10 & 51.6 & 10 &  49.7 & 10 &  82.8 & 10 & 120.9 &  39.6 \\ \hline  
10 & 49.5 &  0 &  73.5 &  0 & 110.4 &  0 & 125.3 &  12.5 \\         
 0 & 58.6 & 10 &  32.1 &  0 & 122.7 &  0 & 172.2 &  25.2 \\         
 1 & 59.6 &  1 &  67.5 & 10 &  75.2 &  1 & 130.9 &  32.0 \\         
 1 & 50.6 &  1 &  60.0 &  1 &  93.0 & 10 & 116.5 &  20.6 \\ \hline  
50 & 47.3 &  0 &  74.0 &  0 & 123.9 &  0 & 139.3 &  80.5 \\         
 0 & 66.8 & 50 &  30.6 &  0 & 140.0 &  0 & 189.1 &  58.6 \\         
 1 & 63.4 &  1 &  70.4 & 50 &  71.6 &  1 & 140.0 &  92.9 \\         
 1 & 50.5 &  1 &  61.6 &  1 &  96.6 & 50 & 112.6 & 113.8 \\         
\hline
\end{tabular}
  \end{center}
  \caption{$\chi^2$ for jet experiments with various weights}
  \label{table:table1}
\end{table}
\endgroup

We will carry out these studies explicitly for the case of the four inclusive 
jet data sets from the Tevatron: CDF Run I, D0 Run I, CDF Run II, D0 Run II.  
Of these, only the Run I sets were included in the CTEQ6--CTEQ6.6 analyses.  
It is well known that the Run I data had a substantial impact on the 
determination of the gluon distribution at large $x$.  It will be interesting 
to see whether the Run II data, which are based on a much larger integrated 
luminosity, provide significant new constraints.  It will also be interesting 
to see whether the Run I data still play a significant role after the 
higher-statistics Run II data have been included.
It will further be interesting to see whether the data from Run II pull the fit 
in the same directions as Run I, or if there is tension between the implications 
of the old and new data sets.  We can similarly ask about possible tension between 
the CDF and D0 data sets.  These are questions that have been raised at a number 
of workshops, but they have not been approached with the methods we 
describe here.

The information needed to answer these questions is contained in 
Table~\ref{table:table1}, which shows $\chi^2$ for each of the 4 jet experiments 
obtained by minimizing the total weighted $\chi^2$ under a variety of choices for 
the weights assigned to those experiments.  The weighted $\chi^2$ for the sum of 
all nonjet experiments is shown in the last column, with the no-jets best fit 
value subtracted for convenience.

The question of whether the jet data sets agree with theory according to 
the ``hypothesis testing'' criterion is answered 
by seeing whether the $\chi^2$ for these sets lie within the expected 
statistical range $N \pm \sqrt{2\,N}$, where $N$ is the number of data 
points in the experiment.
\begin{enumerate}
\item
For CDF$_{\mathrm{I}}$, the expected range is 25--41.  The fit with all jet 
weights 1 lies a little outside that range.  This appears to result from 
unusually large fluctuations in a couple of the data points: 
these data cannot be fitted at much better $\chi^2$ using \textit{any} 
plausible smooth function, as is evidenced by the fact that $\chi^2$ drops 
to only 47 when a weight of 50 is assigned to this experiment.
(The purpose of including fits with a such a large weight in 
Table~\ref{table:table1} is exactly to obtain this kind of information.)
Unlike the other jet experiments, CDF$_{\mathrm{I}}$ has data only in the 
central rapidity region.   It is therefore less sensitive to the gluon 
distribution than the others---in spite of its historic importance in changing 
the view of the gluon at large $x$!  The range of $\chi^2$ for this experiment 
over the entire series of fits shown in the table is quite small, and it
therefore has rather little influence on the contemporary global fit.

\item
For D0$_{\mathrm{I}}$, the expected range is 77--103.  The best-fit $\chi^2$
in the fit with all jet weights equal to 1 is 59, dropping to 32 at weight 10 for this
experiment. If only the Run I data with weight 1 are included, we obtain $\chi^2 = 47$
with the new gluon parametrization, $68$ for CTEQ6.5M, $99$ for CTEQ6.6M,
$124$ and $138$ for restricted gluon parametrizations shown in Tables
2 and 3. Thus this experiment is certainly consistent with the
theory, although the unexpectedly large range of variations in
$\chi^2$ (despite the similarity in the explored PDF parametrizations) 
is suggestive of pronounced non-Gaussian behavior of the systematic errors 
for this data set.  The fact that fits to these data can be obtained with 
with $\chi^2$ so much smaller than the number of points also suggests 
that there is something peculiar about the errors.
(The correlated systematic errors 
for this experiment were published only as a single covariance matrix, 
rather than being broken out as individual shifts associated with each specific 
source of systematic error, whose magnitudes can be directly examined for 
plausibility.  Systematic errors given in this form could nevertheless be 
analyzed using principal component analysis \cite{PavelCovariance}, but we have 
not yet carried this out.)

\item
For CDF$_{\mathrm{II}}$, the expected range is  60--84.  The fit gives 
$\chi^2 = 88$ with all jet weights equal to 1, dropping to 75 for weight 10; 
which implies that these data are consistent with theory.

\item
For D0$_{\mathrm{II}}$, the expected range is  95--125.  The fit gives 
$\chi^2=121$ with all jet weights equal to 1, dropping to 116 for weight 10; 
again eminently consistent with theory.
\end{enumerate}

The question of whether the jet data sets are consistent with the rest of the 
data in the global analysis can be addressed by observing the 
increase in $\chi^2$ for the nonjet data that occurs when the fit is adjusted 
to accommodate the jet data.  Table~\ref{table:table1} shows that $\chi^2$ for 
the nonjet data is forced to increase by only 9.6 to accommodate the 4 jet
experiments at weight 1, and only by 39.6 to accommodate them at weight 10.
In our previous studies of these data for CTEQ6.6, we estimated that an increase 
of $\Delta \chi^2 = 100$ could be tolerated at the 90\% confidence level, 
so the jet experiments appear consistent with the nonjet data. 
Note that we take the ``hypothesis testing'' point of view of requiring that 
$\chi^2$/N be 
acceptable for all of the experiments, rather than the more stringent 
``parameter fitting'' ($\Delta\chi^2 = 1$) point of view for estimating 
the uncertainty limits \cite{CollinsPumplin}.

The question of whether the four inclusive jet experiments are 
consistent with each other in the fit can be studied by looking 
at how increasing the weight for some of them affects the $\chi^2$ for 
the others. From Table~\ref{table:table1}, we observe the following:
\newline
\noindent$\bullet$ \textit{The two Run II experiments are fairly consistent 
with each other,} since for example when 
CDF$_{\mathrm{II}}$ is assigned weight 10, its $\chi^2$ is not strongly 
dependent on whether D0$_{\mathrm{II}}$ is assigned weight 1 or 10; 
and similarly when 
D0$_{\mathrm{II}}$ is assigned weight 10, its $\chi^2$ is not strongly 
dependent on whether CDF$_{\mathrm{II}}$ is assigned weight 1 or 10.  
However, in each case there is a small increase in $\chi^2$ for one 
of the experiments when the weight for the other is increased, which 
suggests a bit of tension between them.  That is 
in fact the case, as can be seen clearly using a new and more powerful 
method of analysis that is discussed in a separate publication \cite{Measuring}.
\newline
\noindent$\bullet$ \emph{The consistency between Run I and Run II measurements 
is ambiguous.} 
If the Run II experiments are assigned
weight 10, then raising the weight for Run I data from 1 to 10 improves the fits 
to Run I as it must, while making very little change in the $\chi^2$ for the 
Run II and nonjet experiments.  This suggests that Run I and Run II data 
are rather compatible with one another.
On the other hand, if instead the Run I experiments are assigned weight 10, 
then raising the weight for Run II data from 1 to 10 (which improves the 
fits to Run II dramatically) raises $\chi^2$ for D0$_{\mathrm{I}}$ 
from 38.6 to 49.7.  An increase of this magnitude suggests tension 
between Run I and Run II---and, indeed,  
the Run I and Run II experiments prefer somewhat different shapes of
the gluon PDF, as will be shown in Sec.\ \ref{sec:GluonUncertaintyResults}.
Yet the statistical significance of this level of 
disagreement cannot be established firmly, given the abnormally 
large variations in $\chi^2$ for D0$_{\mathrm{I}}$ that are observed 
for otherwise very similar
fits. This may be related to the same details of the 
systematic error treatment in D0$_{\mathrm{I}}$ that allows 
$\chi^2$/N to become very small for that experiment.
We keep the Run I data in our final global fit. The fact that the Run I and
Run II experiments are at somewhat different $\sqrt{s}$ values
($1.80 \, \mathrm{TeV}$ vs. $1.96 \, \mathrm{TeV}$) might possibly supply some
useful physics constraint.  Also, D0$_{\mathrm{I}}$ extends to higher rapidity than
either of the Run II data sets.  The effect of this choice will be studied in
Sec.\ \ref{sec:GluonUncertaintyResults} by examining the effect of instead 
dropping the Run I data.

Finally, let us address the question of whether the Run II jet data can be 
expected to reduce the PDF uncertainty.  Table~\ref{table:table1} shows that 
the fit with weight 1 for both Run I experiments and weight 0 for both Run II 
experiments has $\chi^2 = 106$ and $138$ for the two Run II experiments.
Trying each of the 44 eigenvector uncertainty sets of CTEQ6.6, 
we obtain extreme $\chi^2$ values 
of 119 for CDF$_{\mathrm{II}}$ and 140 for D0$_\mathrm{II}$.  None of these 
values indicate a drastically bad fit, so no great reduction in the PDF uncertainty 
can result from including the new jet data.  However, some of the values both for 
CDF$_{\mathrm{II}}$ and  D0$_{\mathrm{II}}$ are sufficiently larger than the values 
shown in Table~\ref{table:table1},
that \emph{we can expect a small reduction in the PDF uncertainty as a result 
of including the new data.} That reduction in uncertainty is examined directly 
in Sec.\ \ref{subsec:LagrangeMultiplier}.
\subsection{Fits with restricted gluon parametrizations}
\label{subsec:RestrictedGluon}

As discussed in Sec.\ \ref{sec:parametrizing}, it is important to 
use a sufficiently flexible parametrization for the input gluon 
distribution.  The following studies demonstrate how an 
inadequate parametrization can be exposed by the weighting method.

If we restrict 
the parametrization (\ref{eq:GluonParametrization}) by setting 
$a_4 \! = \! a_5 \! = \! 0$, we obtain the results shown in 
Table~\ref{table:table2}.
With that restriction, the fit to data without jets is still very 
good: $\chi^2$ is higher by only 2 units.  But the fit to these nonjet 
data becomes very bad when the jet weights are raised to 10 or more; 
while for smaller jet 
weights, the fits to the jet experiments are much worse than the 
corresponding fits of Table~\ref{table:table1}.  If this 
simplified parametrization had been used, the jet data would have 
mistakenly appeared to be inconsistent with the rest of the data.

\begin{table}[!hbt]
\squeezetable
  \begin{center}
\begin{tabular}{||rr|rr|rr|rr||r||}
\hline
\multicolumn{2}{||c|}{CDF$_{\mathrm{I}}$  N=33} & 
\multicolumn{2}{|c|}{D0$_{\mathrm{I}}$ N=90} &
\multicolumn{2}{|c|}{CDF$_{\mathrm{II}}$ N=72} & 
\multicolumn{2}{|c||}{D0$_{\mathrm{II}}$ N=110} &
$\Delta \chi^2$ \\ 
Wt & $\chi^{2}$ & Wt & $\chi^{2}$ & Wt & $\chi^{2}$ & Wt & $\chi^{2}$ & nonjet \\ 
\hline
 0 & 55.8 &  0 & 145.9 &  0 & 120.6 &  0 & 155.2 &   2.0 \\        
 1 & 53.2 &  1 & 124.0 &  0 & 118.2 &  0 & 148.6 &   7.7 \\        
 0 & 58.6 &  0 & 121.3 &  1 &  98.1 &  1 & 137.8 &  16.8 \\        
 1 & 54.5 &  1 & 108.8 &  1 &  95.5 &  1 & 134.2 &  25.8 \\        
10 & 54.1 & 10 &  75.7 &  0 & 142.0 &  0 & 152.1 & 184.3 \\ \hline 
10 & 51.9 & 10 &  74.0 &  1 & 101.6 &  1 & 134.6 & 185.5 \\        
 0 & 67.1 &  0 &  75.3 & 10 &  77.3 & 10 & 126.1 & 114.6 \\        
 1 & 60.3 &  1 &  74.1 & 10 &  77.1 & 10 & 125.8 & 119.3 \\        
10 & 51.7 & 10 &  64.5 & 10 &  76.2 & 10 & 126.1 & 204.3 \\        
\hline
\end{tabular}
  \end{center}
  \caption{Fits to jet experiments with various weights, using 
a restricted gluon parametrization.}
  \label{table:table2}
\end{table}

A different simplified gluon parametrization 
\begin{equation}
g(x) = a_0 \, x^{a_1} \, (1 - x)^{a_2} \, 
       (1 \, + \, a_3 x) \; ,
  \label{eq:ZeusForm}
\end{equation}
which has been used in studies at HERA \cite{Zeus} 
(at a somewhat higher $\mu_0$),
has even worse behavior, as is show in Table~\ref{table:table3}.  
For here, $\chi^2$ for the nonjet data rises by 91.4 when the jet data 
are included at weight 1; and that weight is not even large enough to obtain 
good fits to the jet data.  It is perhaps not surprising that the form 
(\ref{eq:ZeusForm}) is inadequate, because 
the coefficient $a_0$ of the leading behavior $x^{a_1}$ at $x \to 0$ and 
the coefficient $a_0 \, (1 + a_3)$ of the leading behavior $(1 - x)^{a_2}$ 
at $x \to 1$ might have very different magnitudes, since those limits are 
governed by unrelated physics.  Hence the limiting behaviors 
might require $1 + a_3$ to be very large or very small, in which case the 
linear approximation $1 + a_3 x$ provided by 
Eq.~(\ref{eq:ZeusForm}) would have to cover a large range of variation,
for which it might be a worse approximation than the exponential form 
in (\ref{eq:GluonParametrization}).

\begin{table}[!hbt]
\squeezetable
  \begin{center}
\begin{tabular}{||rr|rr|rr|rr||r||}
\hline
\multicolumn{2}{||c|}{CDF$_{\mathrm{I}}$  N=33} & 
\multicolumn{2}{|c|}{D0$_{\mathrm{I}}$ N=90} &
\multicolumn{2}{|c|}{CDF$_{\mathrm{II}}$ N=72} & 
\multicolumn{2}{|c||}{D0$_{\mathrm{II}}$ N=110} &
$\Delta \chi^2$ \\ 
Wt & $\chi^{2}$ & Wt & $\chi^{2}$ & Wt & $\chi^{2}$ & Wt & $\chi^{2}$ 
& nonjet\\ 
\hline
 0 & 57.8 &  0 & 175.4 &  0 & 139.5 &  0 & 180.9 &   4.3 \\        
 1 & 54.5 &  1 & 137.9 &  0 & 148.5 &  0 & 165.8 &  91.4 \\        
 0 & 64.0 &  0 & 143.7 &  1 & 102.4 &  1 & 151.4 &  27.8 \\        
 1 & 53.9 &  1 & 128.4 &  1 & 105.4 &  1 & 143.0 & 110.7 \\        
10 & 56.3 & 10 &  74.2 &  0 & 174.7 &  0 & 166.2 & 235.1 \\ \hline 
10 & 53.5 & 10 &  75.6 &  1 & 115.8 &  1 & 139.6 & 299.2 \\        
 0 & 79.3 &  0 &  79.0 & 10 &  75.8 & 10 & 131.8 & 156.9 \\        
 1 & 68.5 &  1 &  77.8 & 10 &  75.2 & 10 & 131.7 & 163.9 \\        
10 & 54.4 & 10 &  67.7 & 10 &  72.8 & 10 & 131.8 & 259.0 \\        
\hline
\end{tabular}
  \end{center}
  \caption{$\chi^2$ for jet experiments with various weights, using 
the restricted gluon parametrization (\ref{eq:ZeusForm}).}
  \label{table:table3}
\end{table}

\begin{figure}[tbh]
\begin{center}
 \resizebox*{0.42\textwidth}{!}{
 \includegraphics{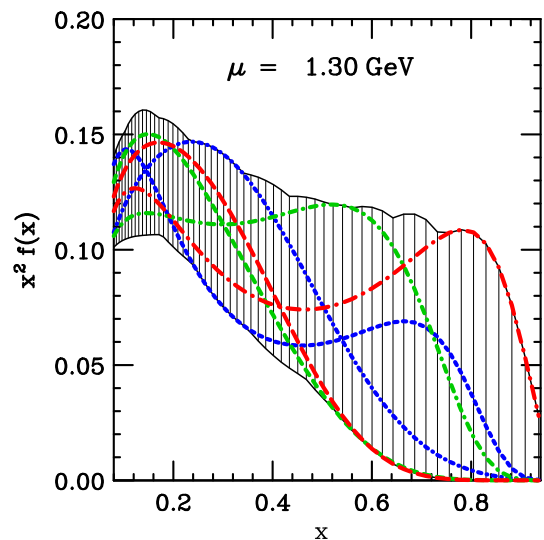}
 }
\end{center}
 \caption{Gluon uncertainty range by LM method, and some of the 
specific fits that define the limits.
}
 \label{fig:figThree}
\end{figure}

\section{Uncertainty of the gluon distribution: Comparison of methods}
\label{sec:GluonUncertaintyMethods}

In this section, we discuss various methods to determine the uncertainty 
of parton distributions.  
We focus on the gluon distribution at large $x$ 
because that is the primary aspect of the global analysis that is influenced 
by the jet experiments.
Because the uncertainty in the gluon
is large, it serves as a strong test of the methods used to estimate 
uncertainties.

Within the usual context of our global analysis \cite{cteq606165}, 
parton distribution shape parameters that minimize an effective
weighted $\chi^2$ function define the ``best fit''. 
All parton distributions defined by the other choices of the parameters are deemed 
acceptable (and delineate the region of the PDF uncertainty allowed by
the analysis) if they produce a value of $\chi^2$ that exceeds the
minimum value by no more than a given tolerance value $\Delta \chi^2$ 
(i.e., $\chi^2\leq \min(\chi^2) + \Delta \chi^2$). Appropriate weights and the
tolerance criterion must be chosen to ensure that all of the accepted fits 
provide adequate descriptions of every data set.  In the present case, we 
estimate that $\Delta \chi^2 = 100$ provides an approximately 90\% confidence
limit for all experiments included in the fit. 

\begin{figure}[tbh]
\begin{center}
 \resizebox*{0.42\textwidth}{!}{
 \includegraphics{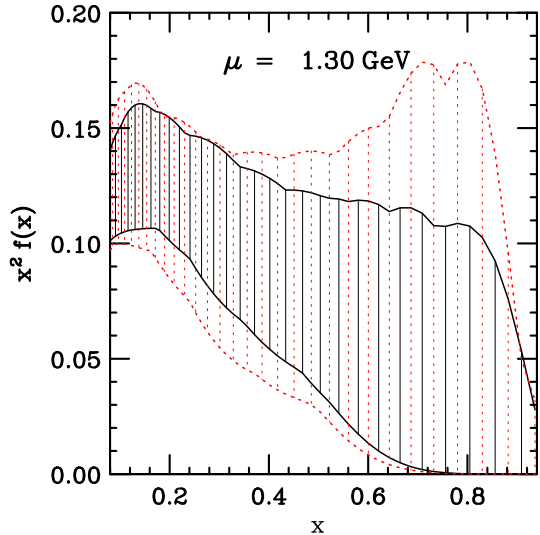}
 }
\end{center}
 \caption{Gluon uncertainty range by LM method: smaller region same as 
in Fig.\ \ref{fig:figThree}; larger region = without Run II jet data.
}
 \label{fig:figFour}
\end{figure}

\subsection{Lagrange Multiplier method}
\label{subsec:LagrangeMultiplier}
The uncertainty of the gluon distribution can be found in a straightforward way 
by the Lagrange Multiplier (LM) method \cite{Lagrange}:  at any given value of 
$x$, a term $\lambda \, g(x,\mu_0)$ is added to the $\chi^2$ function that is 
minimized by varying the fitting parameters.  The parameter $\lambda$ is 
adjusted to make the increase in $\chi^2$ above its minimum value equal to 
$\Delta \chi^2$.  This yields two allowed PDF sets (one from positive $\lambda$ 
and one from negative) that provide the minimum and maximum $g(x,\mu_0)$.  
The procedure is carried out at a number of $x$ values to map out the extremes 
of the uncertainty range.

Results for the gluon uncertainty obtained in this way are shown in 
Fig.~\ref{fig:figThree}, together with some of the specific curves that 
produced the envelope of extremes.  
The shapes that provide the extremes do not violate any strong 
intuition, although those showing a peaked structure in $x^2 \, g(x,\mu_0)$ 
at large $x$ might not be expected {\it a priori}.
(Still larger uncertainties might be found if more fine structure were 
allowed by the parametrization; but sharp structures in $x$ are not physically 
expected, and their effect would tend to go away at higher scales 
through the smoothing character of DGLAP evolution.)

It is natural to ask if the extensive new jet data from Run II reduce 
the gluon uncertainty.  To answer that question, Fig.~\ref{fig:figFour} 
compares the uncertainty range from Fig.\ \ref{fig:figThree} with the 
uncertainty range obtained by a similar Lagrange Multiplier calculation 
with the Run II data removed from the fit.  
\textit{One sees that the Run II data somewhat reduce the gluon uncertainty 
at large $x$.}

\subsection{Quartic penalties}
\label{subsec:ModifiedWeights}
A PDF set that deviates from the minimum $\chi^2$ 
by an amount $\Delta \chi^2 = 100$ usually provides an 
acceptable fit to all experiments and thus cannot be ruled out as a valid 
possibility within the uncertainty range according to the conservative 
``hypothesis testing'' criterion.  
But if the increase in $\chi^2$ is not 
spread widely over the $\sim$3000 data points, but rather is concentrated in 
one or two experiments, or in any small subset of the data points, it may be 
an unacceptable fit.  This is found to happen 
for some of the 
extreme gluon distributions obtained in Sec.\ \ref{subsec:LagrangeMultiplier}, 
because only the inclusive jet experiments are sensitive to the gluon 
distribution at large $x$.

To avoid this problem, we could increase the weight for the jet experiments 
in the total $\chi^2$ by trial and error.  But we find it simpler and more 
effective to add a penalty to $\chi^2$ that is proportional to 
$(\chi^2/N)^4$ for each of the jet experiments, 
in order to force the final fit to agree acceptably with each of those
experiments, without introducing much change in the central fit.  
With this change in the definition of the 
weighted $\chi^2$ that is minimized, we can continue to use our established 
calculational 
tools. (An alternative method used by MSTW \cite{MSTW2008} is to 
abandon a fixed $\Delta \chi^2$ and instead to set the maximum allowed 
displacement along each eigenvector direction independently, by monitoring 
the quality of fit to each of the data sets along that direction.)
The quartic form for the 
penalty adds little to $\chi^2$ except near the boundary, so it does not 
significantly alter our $\Delta \chi^2 = 100$ tolerance estimate.

These ``quartic penalties'' are included in all subsequent fits
in this paper.  Our final uncertainty for the gluon distribution is therefore 
appreciably smaller than what is shown in the preliminary study of 
Figs.\ \ref{fig:figThree} and \ref{fig:figFour}.

\begin{figure}[tbh]
\begin{center}
 \resizebox*{0.42\textwidth}{!}{
 \includegraphics{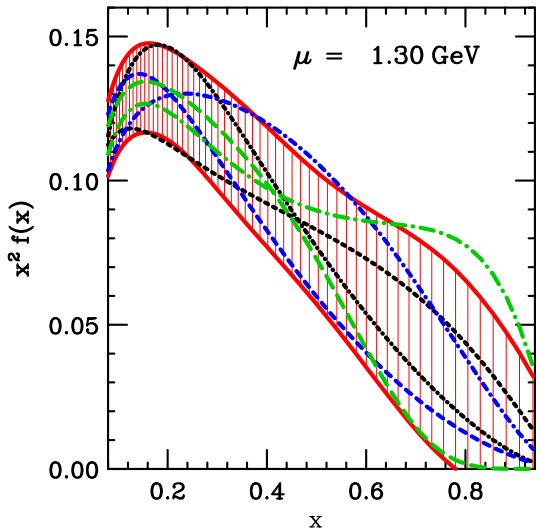}
 }
\end{center}
 \caption{Gluon uncertainty by Hessian method, compared to 
extremes at $x=0.20, 0.55, 0.80$ by LM method.
}
 \label{fig:figFive}
\end{figure}

\subsection{Hessian eigenvector method}
\label{subsec:Hessian}
In addition to the LM method, the other 
standard technique for estimating PDF uncertainties is the Hessian 
eigenvector method \cite{Hessian}.  That method works as follows. The first 
derivatives of $\chi^2$ with respect to the fitting parameters are zero 
at the minimum, so in the neighborhood of the minimum, $\chi^2$ can be 
approximated by Taylor series as a quadratic form in the fitting parameters.  
The coefficients of that quadratic form are the Hessian matrix, 
which is the matrix of second derivatives of $\chi^2$ with respect to the 
fitting parameters.
The eigenvectors of the Hessian matrix can be used to define eigenvector PDF 
sets that characterize the allowed uncertainty range.  The uncertainty of any 
prediction is calculated by computing the deviation from the best fit along 
each eigenvector direction, and adding those deviations in quadrature separately 
for the positive and negative deviations.  The gluon uncertainty calculated this 
way is shown in Fig.\ \ref{fig:figFive}, together with extremes 
calculated by LM at $x=0.2$, $0.5$, and $0.8$.  The agreement between the 
two methods is seen to be quite good, although a slightly larger upper 
limit is found at $x=0.8$ by the LM method, which is not subject to the 
quadratic approximation.
The eigenvector method is of course much more convenient to use than LM,
because the LM method requires tuned fittings of the 
Lagrange Multiplier parameter for every extremum point that is desired.
So it is comforting to see this agreement.

\subsection{Choice of eigenvectors}
\label{subsec:ChoiceOfEigenvectors}

The eigenvectors of the Hessian matrix can be thought of as a choice of basis 
vectors that define new fitting parameters $z_i$ for which
\begin{equation}
\chi^2 \, = \, \chi^2_{\mathrm{min}} \, + \, \sum_{i=1}^N z_i^{\, 2}
\, + \, {\cal O}(z^3) \; .
\label{eq:chidiag}
\end{equation}
The choice of these eigenvectors is not unique, because
the form (\ref{eq:chidiag}) is preserved by any further orthogonal 
transformation of the coordinates $\{z_i\}$.  In the approximation that 
$\chi^2$ is a quadratic function of the shape parameters which parametrize 
PDFs at $\mu_0$, such a transformation would not affect the calculation of 
the uncertainty. 

\begin{figure*}[tbh]
\begin{center}
 \resizebox*{\textwidth}{!}{
\mbox{
\resizebox{0.38\textwidth}{!}{
\includegraphics[clip=true,scale=1.0]{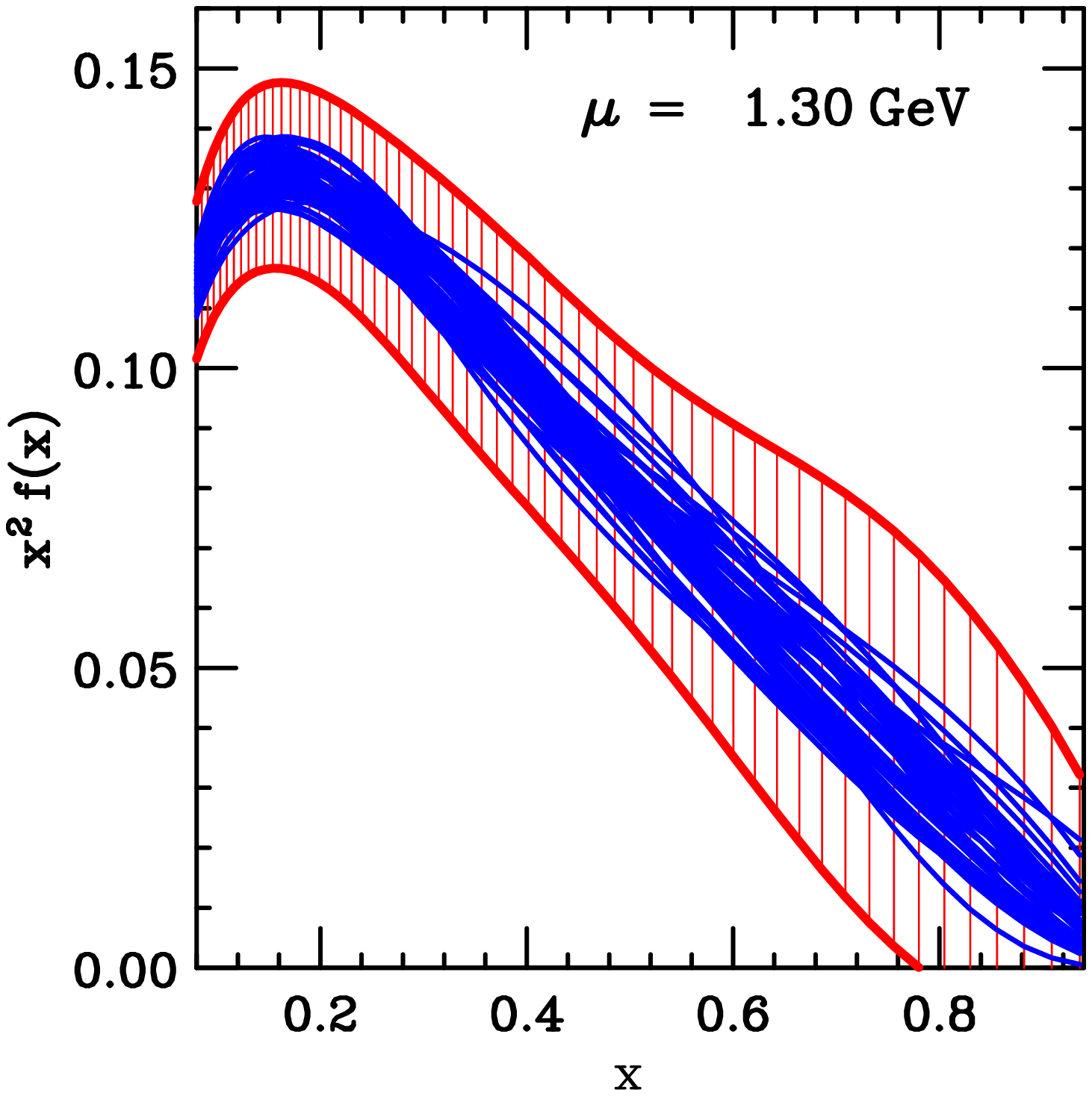}}
\hfill
\resizebox{0.38\textwidth}{!}{
\includegraphics[clip=true,scale=1.0]{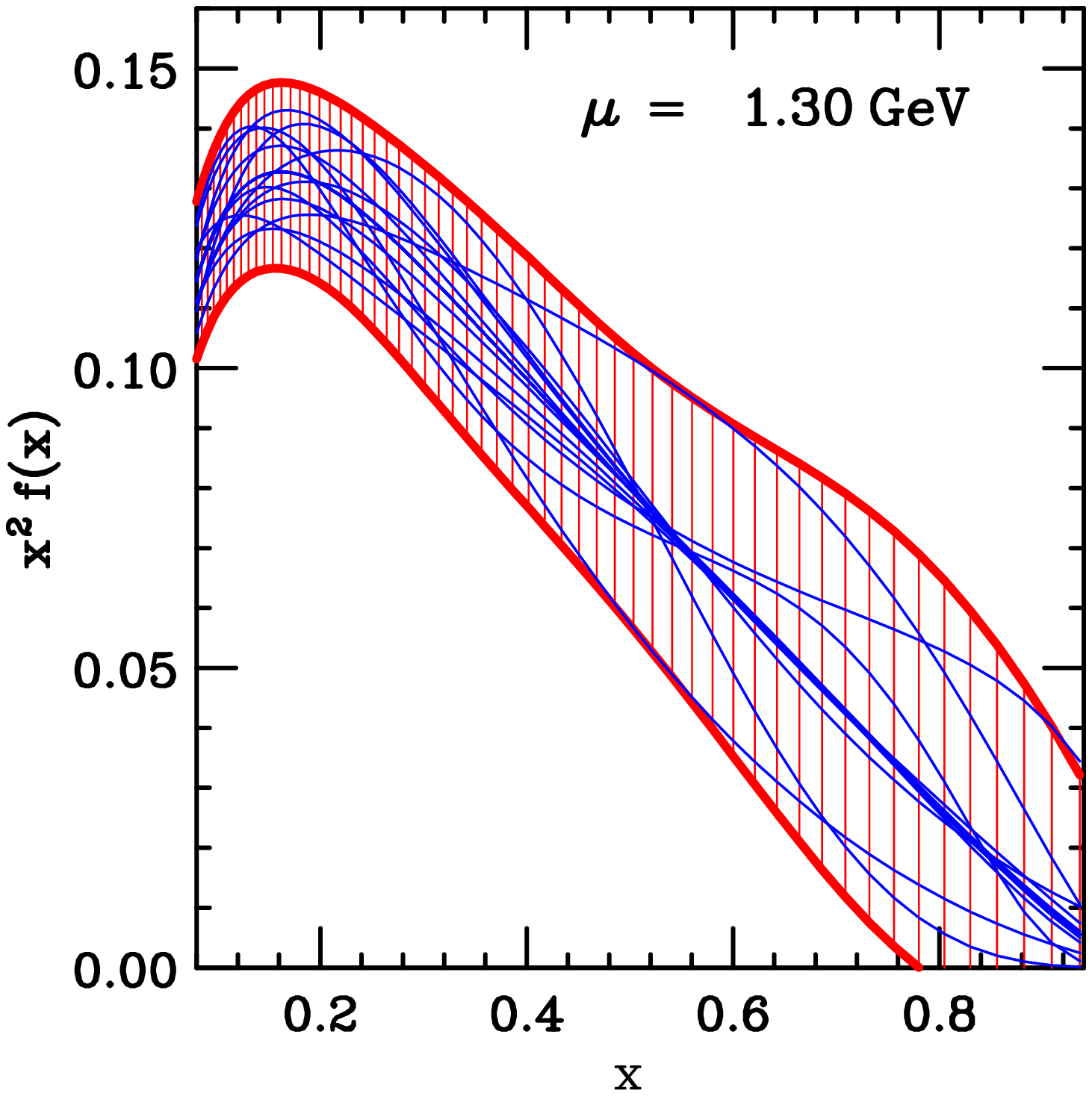}}
}
 }
\end{center}
 \caption{Gluon distributions and uncertainties in CT09 (red) and 
the eigenvector contributions to them.  Left: eigenvectors by 
traditional method; Right, eigenvectors by ``rediagonalization''
method based on diagonalizing $g(\mu_0,x)$ at $x=0.55$.
}
 \label{fig:figSix}
\end{figure*}

The freedom to make an additional orthogonal transformation may offer 
the possibility to reduce the number of eigenvectors that are needed 
to effectively describe the uncertainty of a particular quantity of 
interest.  One possible way to attempt this is to diagonalize the 
parameter dependence of that quantity, using a procedure that is 
sketched in the Appendix and described explicitly in \cite{Measuring}.

An example of this is shown in Fig.~\ref{fig:figSix}, which shows the gluon 
uncertainty calculated by the eigenvector method, together with the 48 extreme 
eigenvector sets (positive and negative directions along each of the 24 
eigenvectors).  In the left panel, the eigenvectors are defined in the 
traditional way as eigenvectors of the 
Hessian.  Note that many eigenvectors contribute to the uncertainty at each 
value of $x$.  
(A common method to make a quick estimate of uncertainty 
is simply to look at the extremes over the eigenvector sets, without adding 
the individual contributions in quadrature.  That can easily underestimate 
the uncertainty by a factor of two or more, as seen here.)

In the right panel of Fig.~\ref{fig:figSix}, the eigenvectors are defined by 
choosing $G = g(0.55,\mu_0)$ in Eq.~(A3) of the Appendix.
Note that close to $x=0.55$, almost all of the uncertainty comes from just 
one pair of eigenvector sets.  In CTEQ6.1, it happened by convenient 
accident that most of the uncertainty in the gluon distribution was embodied 
in a single eigenvector set.  By ``rediagonalizing'' the Hessian matrix, this 
type of simplicity can be gained in other situations; though as seen in 
Fig.~\ref{fig:figSix} it may take more than one eigenvector direction to span 
the important variations.

A rediagonalization based on the second-derivative matrix, such as 
the one carried 
out here, is not necessarily the best way to choose the new eigenvector 
directions, since there is no theorem to guarantee that it will result in only 
a few dominant coefficients.  For example in this particular case it might have 
worked better to ignore the second derivatives, and instead to simply choose the 
first new eigenvector direction along the gradient direction for, say $g(0.5)$ 
in the 24-dimensional space; then the second eigenvector could be chosen along 
the gradient direction for, say $g(0.8)$ in the 23-dimensional subspace that 
is orthogonal to the first eigenvector, etc.  
In any case, the option of redefining 
the eigenvector directions to simplify the description of uncertainties in other 
physics analyses shows promise for further study.

\begin{figure}[tbh]
\begin{center}
 \resizebox*{0.42\textwidth}{!}{
 \includegraphics{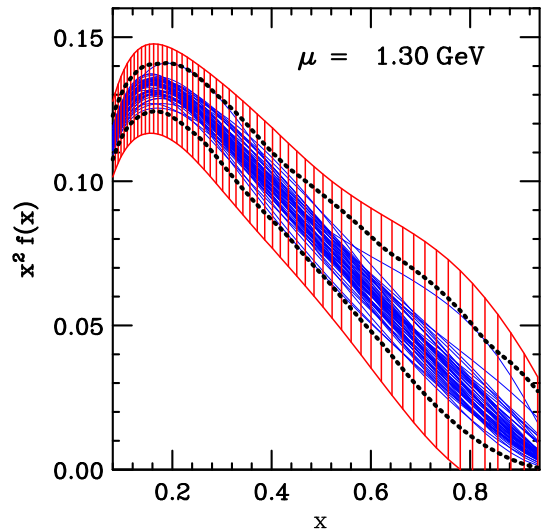}
 }
\end{center}
 \caption{Gluon uncertainty from 50 random PDF sets with 
$\Delta\chi^2=100$; envelope of 500 such random sets (dotted);
full uncertainty range from Hessian method (shaded region).
}
 \label{fig:figSeven}
\end{figure}

\subsection{Random PDF sets}
\label{subsec:RandomPDF}
Another possible way to characterize the uncertainties would be to generate a 
random collection of PDF sets that lie inside or at the edge of the acceptable 
range $\chi^2 \, \le \, \chi^{\, 2}_{\mathrm{min}} + \Delta\chi^2$.  
(In the quadratic approximation, this would correspond to a sphere in the 
N-dimensional hyperspace spanned by $\{z_i\}$.)
For example,
a set at the edge can be constructed by generating a random unit vector in the 
$N$-dimensional parameter space using the eigenvectors as basis vectors,  
and moving away from the minimum point
in that direction until $\chi^2$ has increased by the tolerance $\Delta\chi^2$.
The envelope of results obtained from 500 PDF sets obtained this way is shown 
in Fig.~\ref{fig:figSeven}, together with 50 of the individual results. 
Also shown is the uncertainty obtained by the Hessian method.
\textit{We see that the envelope of the random sets covers a much smaller range 
than the full uncertainty---even though 
every one of the 500 sets is at the upper limit for $\chi^2$.}
This is not surprising, since the extreme $g(x)$ at any given $x$ can 
be thought of as corresponding to a specific direction in the N-dimensional 
parameter space.  The probability distribution for the component, $z$, of a 
random unit vector along any particular direction in $N$ dimensions can be 
shown to be  
$dP/dz \propto (1 - z^2)^{(N-3)/2}$, which becomes extremely small as $z$ 
approaches its limit of 1.
For example, when $N=24$, the probability for $z > 0.6$ is 
less than 1 in 1000, so the chance of finding a value close to the true 
extreme of 1.0 by random sampling is very small.
This conclusion can be understood qualitatively in a simple way: it is 
unlikely for the direction cosine along any particular direction to lie 
close to its maximum of 1, since there are $N$ random direction cosines 
whose sum of squares must add up to 1.

The point of this exercise is to show that no conveniently small collection
of PDFs that are all acceptable fits to the data can approximately cover the
full uncertainty range.  It is therefore essential to have a well-defined
way to combine the uncertainties associated with the various fits in such
a collection.  In the Hessian method, this is provided by the
rule of adding uncertainties from eigenvector sets in quadrature.  In the 
case of random PDF sets, it would require estimating the uncertainty range 
for a prediction of a quantity $X$ using the dispersion 
$\langle X^2 \rangle - \langle X \rangle^2$ in values calculated from the 
random sets.

The above limitation does not apply to Monte Carlo based sampling methods 
such as NNPDF \cite{NNPDF}, since those methods produce a collection of PDF 
sets that directly samples the space of uncertainties.  Such a collection
naturally includes some PDF sets that are not ``acceptable'' fits to the
input data---e.g., in a collection of 100 Monte Carlo sets, one obviously
expects to find $\sim$10 sets that lie outside of the 90\% confidence
region.  In this approach, the PDF uncertainty for a quantity is obtained 
by simply calculating that quantity for each of the sample PDF sets: the 
distribution of results directly represents the predicted uncertainty range.

\begin{figure*}[tbh]
\begin{center}
 \resizebox*{\textwidth}{!}{
\mbox{
\resizebox{0.38\textwidth}{!}{
\includegraphics[clip=true,scale=1.0]{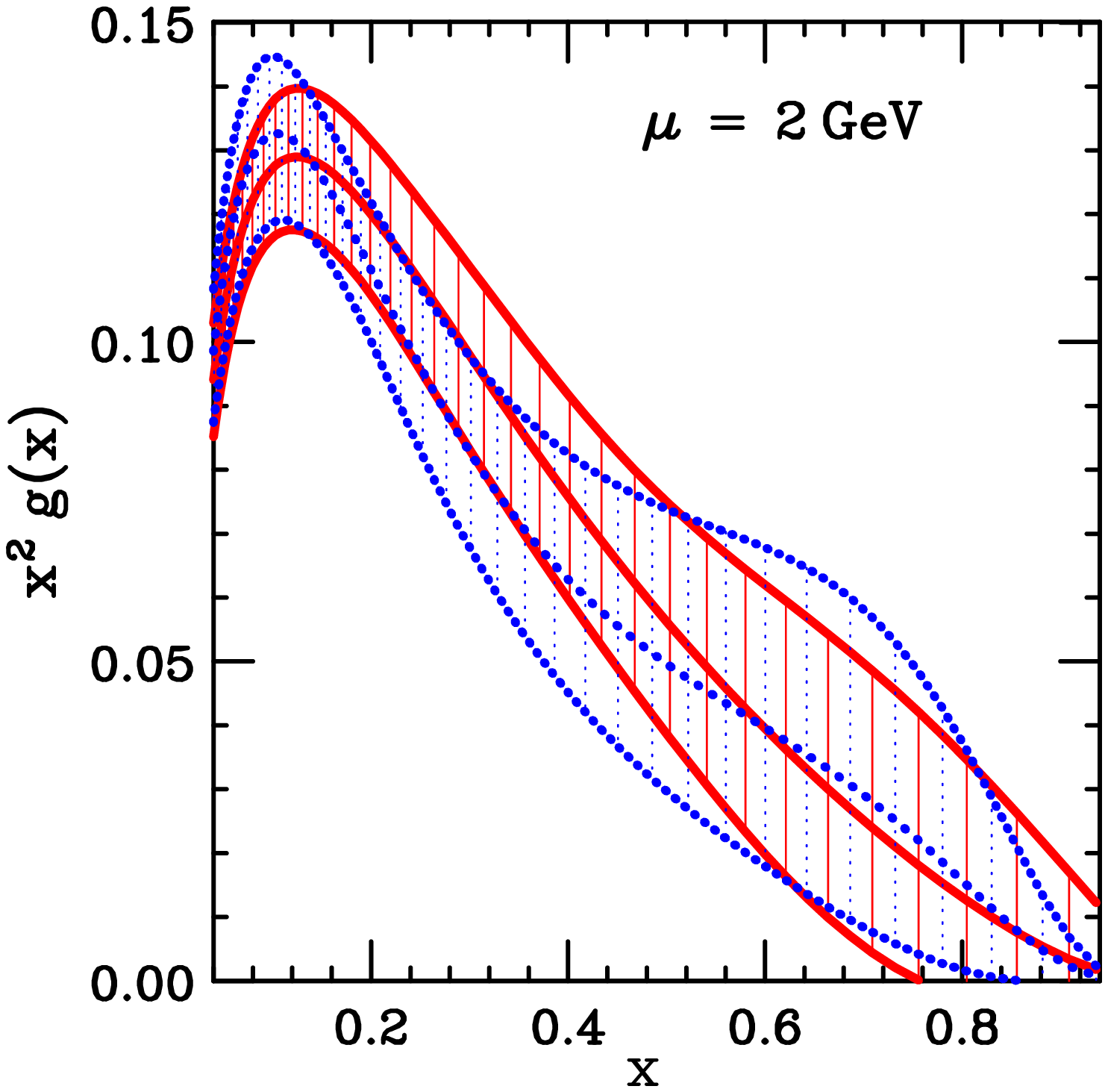}}
\hfill
\resizebox{0.38\textwidth}{!}{
\includegraphics[clip=true,scale=1.0]{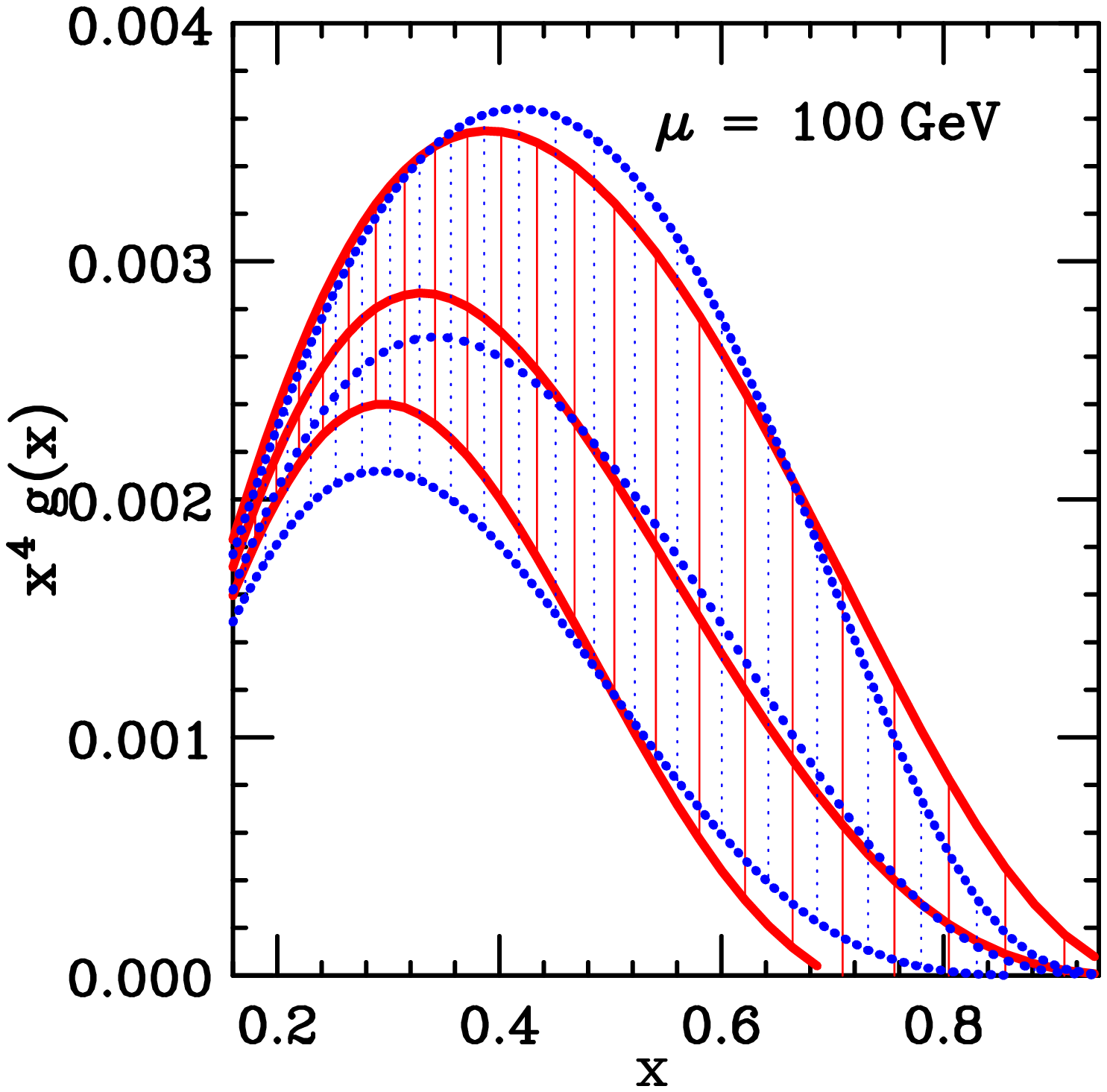}}
}
 }
\end{center}
 \caption{Gluon distributions and uncertainties in CT09 (solid) and 
CTEQ6.6 (dotted).
A stronger weight factor $x^4$ is used in the right-hand 
panel ($\mu = 100 \, \mathrm{GeV}$) to accentuate the large-x behavior.
}
 \label{fig:figEight}
\end{figure*}

\section{Uncertainty of the gluon distribution: Results}
\label{sec:GluonUncertaintyResults}

The CT09 fit discussed in 
Secs.\ \ref{subsec:Hessian} and \ref{subsec:ChoiceOfEigenvectors} is 
our most up-to-date set of parton distributions.
The central gluon fit and its uncertainty are shown in 
Fig.\ \ref{fig:figEight} at scales $\mu = 2 \, \mathrm{GeV}$ 
and $100 \, \mathrm{GeV}$, compared with the previous 
CTEQ6.6 \cite{cteq66} fit.
The uncertainty band has narrowed somewhat as a result of including the new jet 
data and the quartic penalties---except at extremely large $x$, where the more 
flexible gluon parametrization in CT09 has broadened the allowed range.  
There is a strong overlap between the old and new uncertainty bands, and the 
central fit has shifted by an amount that is within or just at the edge 
of those bands.
At a large scale such as $\mu = 100 \, \mathrm{GeV}$, there is rather little
change between the old and new determinations.

Fig.\ \ref{fig:figEight} shows that the CT09 central fit at small scale has a 
featureless behavior at large $x$, in contrast to the mild ``shoulder'' 
structure of CTEQ6.6.  (The appearance of this shoulder is enhanced 
by the factor $x^2$ that multiplies $g(x)$ in the plot to 
emphasize the large $x$ behavior.) Indeed, MSTW \cite{MSTW2008} remark that 
in fitting the new jet data, they no longer need to use their former convoluted 
method of parametrizing the gluon in the DIS scheme and transforming it to 
$\overline{\mathrm{MS}}$.
However, we find that with a properly flexible parametrization, some type of 
shoulder structure is not ruled out---indeed, the original CTEQ6.6 central fit 
for the gluon distribution still lies within our allowed uncertainty range.
In detail, $\chi^2$ for the jet experiments 
(CDF$_{\mathrm{I}}$, D0$_{\mathrm{I}}$, 
CDF$_{\mathrm{II}}$, D0$_{\mathrm{II}}$) are 
$(54,\,59,\,91,\,122)$ in CT09; 
$(52,\,55,\,116,\,121)$ in a fit with the gluon shape identical to CTEQ6.6; and
$(53,\,60,\,97,\,120)$ in a fit using the CTEQ6.6 gluon parametrization with the
parameters refitted.

\begin{figure}[tbh]
\begin{center}
 \resizebox*{0.42\textwidth}{!}{
 \includegraphics{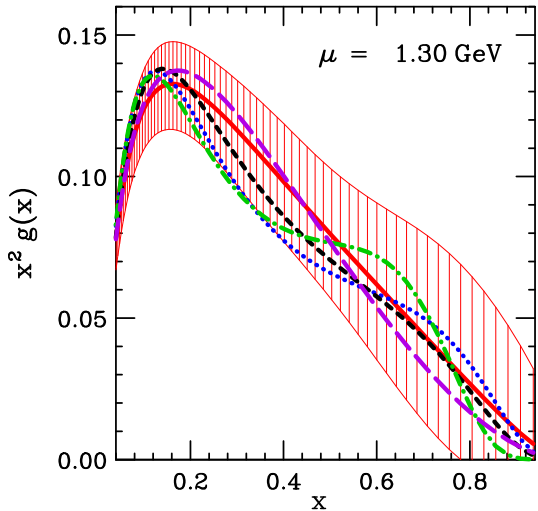}
 }
\end{center}
 \caption{CT09 and variations (see text).}
 \label{fig:figNine}
\end{figure}

The change between CTEQ6.6 and CT09 in the shape of the gluon distribution 
is a consequence of interplay between adding the Run II jet data and 
increasing the flexibility of the gluon parametrization.  This is studied in 
Fig.\ \ref{fig:figNine}.  The solid curve and shaded region are again CT09 and
its uncertainty.  The dotted curve is CTEQ6.6.  The dot-dash curve is the 
result of repeating the CTEQ6.6 fit using the CT09 gluon parametrization.
Note that this increased freedom for the gluon shape enhances the shoulder, 
and does not move the fit closer to 
CT09.  The short-dashed curve is the result of including the Run II data, 
with all other details of the fit being the same as in CTEQ6.6: this 
changes the fit about half way to CT09.  But with the Run II data included,
bringing in the more flexible gluon parametrization now produces the rest 
of the change to CT09.  Finally, the long-dashed curve is a fit that is 
identical to CT09 except for dropping the Run I jet data.  This answers
the question raised earlier regarding the degree of tension between Run I 
and Run II jet data from a practical point of view:  we see that the effect 
of the Run I data on the fit is noticeable but small compared to the other 
uncertainties. 

\begin{figure*}[tbh]
\begin{center}
\mbox{
 \resizebox*{0.40\textwidth}{!}{
\includegraphics[clip=true,scale=1.0]{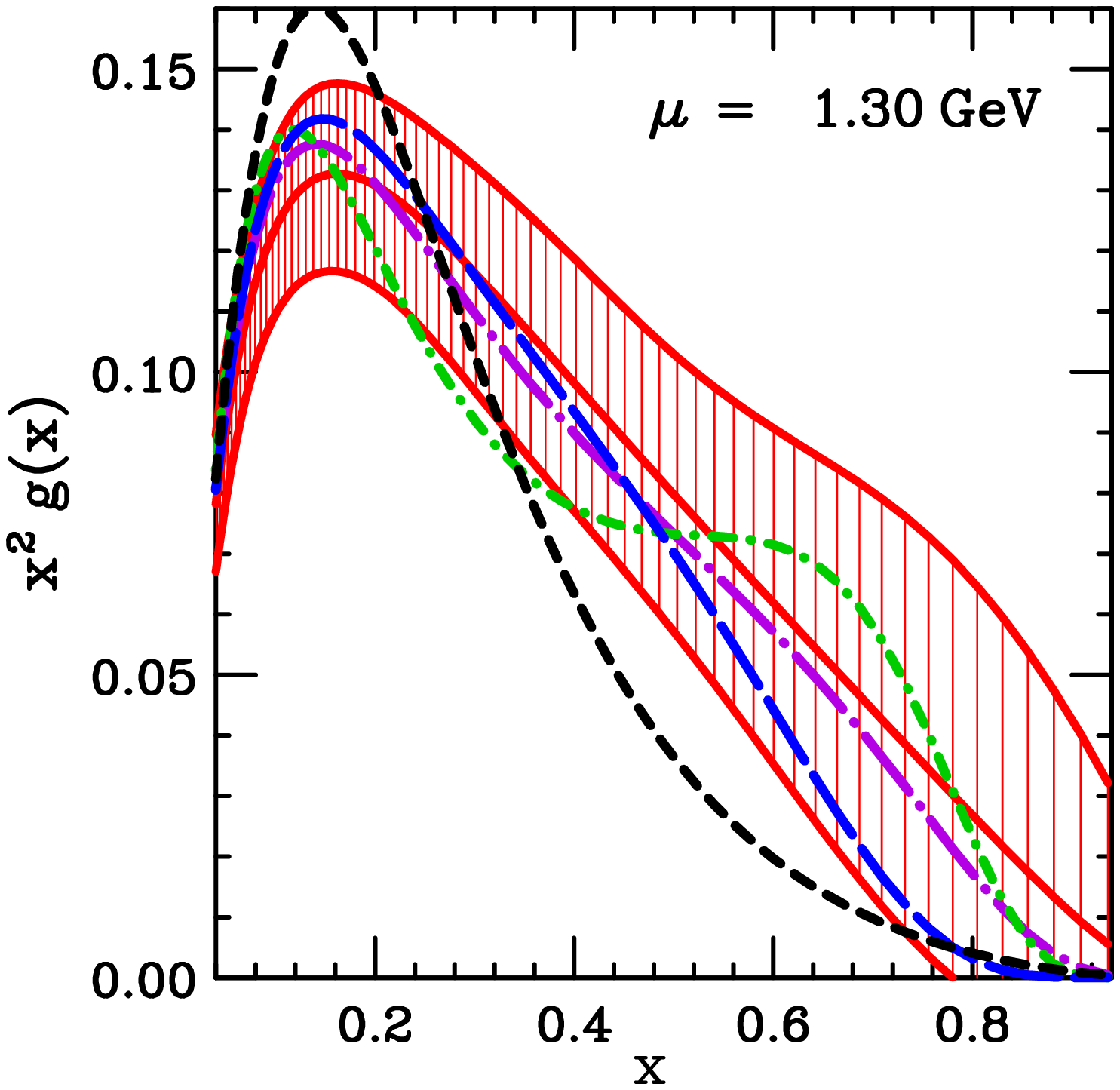}}
\hfill
 \resizebox{0.40\textwidth}{!}{
\includegraphics[clip=true,scale=1.0]{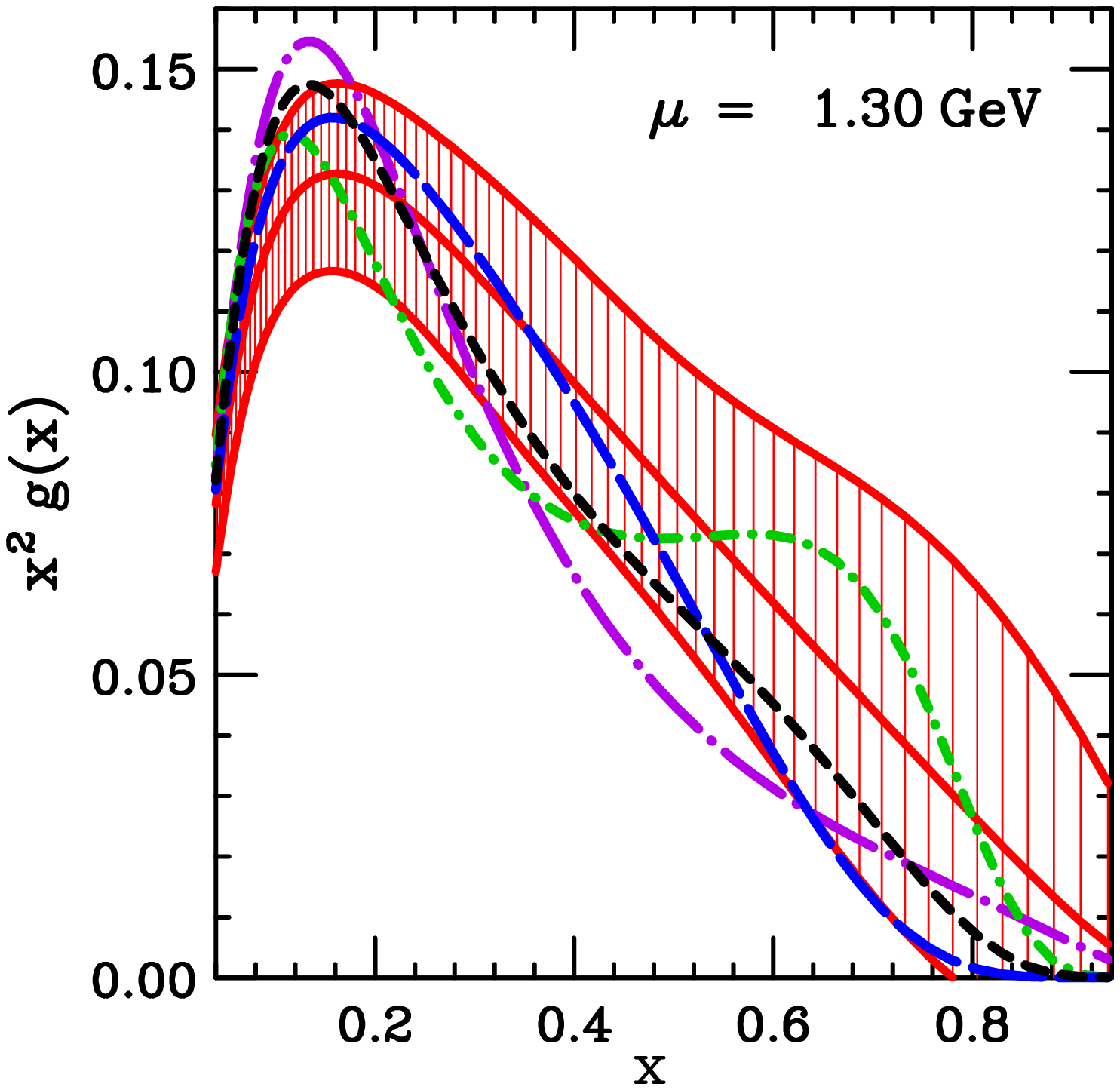}}
}
\end{center}
 \caption{
Fits with various weights 
(CDF$_{\mathrm{I}}$, D0$_{\mathrm{I}}$, 
CDF$_{\mathrm{II}}$, D0$_{\mathrm{II}}$).
Left panel: 
Long dashed dotted = (1,1,1,1), 
Short dashed dotted = (1,1,0,0), 
Long dashed = (0,0,1,1),
Short dashed = (0,0,0,0). 
Right panel:
Long dashed dotted = (1,0,0,0), 
Short dashed doted = (0,1,0,0), 
Long dashed = (0,0,1,0),
Short dashed = (0,0,0,1). 
}
 \label{fig:figTen}
\end{figure*}

It is instructive to examine the preferences of various combinations of 
the four jet data sets in the fit.  This is shown in 
Fig.\ \ref{fig:figTen}.  The solid curve and shaded region are again CT09 in 
both panels.  The other curves were obtained by fits with weight 1 for all 
nonjet experiments, and weights 0 or 1 for each jet experiment as listed 
in the captions.  The four curves in the left panel correspond to the first 
four fits in Table~\ref{table:table1}.  There is a slight difference between
the (1,1,1,1) curve and CT09, because we have chosen to apply somewhat 
larger weights (1.3,1.3,2.1,2.1) to these experiments in CT09.  The fit 
with no input from jet data (0,0,0,0) is substantially lower than any of 
the other fits at large $x$---this is a review of why the first jet data 
made such a strong impact on the gluon determination!
The four curves in the right panel show the preferences of the individual 
jet experiments. 
The D0$_{\mathrm{I}}$ data shows its famous preference for a peak at large 
$x$; though Table~\ref{table:table1} shows that it can be fit with nearly 
as good $\chi^2$ without the peak.
The difference between the CDF$_{\mathrm{II}}$ and D0$_{\mathrm{II}}$ curves 
is comparable to our error estimate, which affirms that our error estimate 
is not overly conservative.  

\begin{figure}[tbh]
\begin{center}
 \resizebox*{0.42\textwidth}{!}{
 \includegraphics{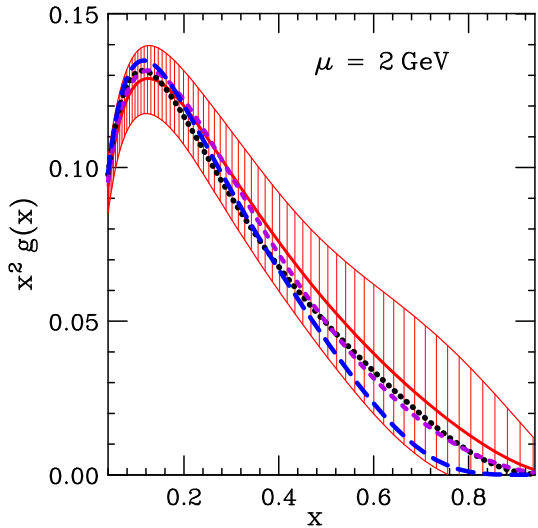}
 }
\end{center}
 \caption{
CT09 and results from some alternative choices (see text).
}
 \label{fig:figEleven}
\end{figure}

Figure \ref{fig:figEleven} explores the consequences of some 
of the choices that were made in producing CT09.  The solid curve and 
shaded region are CT09 itself at scale $2 \, \mathrm{GeV}$.  
We first change the quark masses $m_c = 1.3 \to 1.4 \, \mathrm{GeV}$ and 
$m_b = 4.5 \to 4.75 \, \mathrm{GeV}$,
and change 
$\mu_0 = 1.3 \to 1.4 \, \mathrm{GeV}$ to maintain $\mu_0 = m_c$.  These 
changes are found to have a negligible effect on the gluon distribution:
the change is smaller than the width of the line in the figure.

In our basic fitting procedure \cite{cteq606165} we routinely employ weight 
factors to improve the quality of fit to certain key experiments.  
In particular, weights 
of 1.3 and 2.1 were applied to the Run I and Run II data respectively in CT09, 
and a further contribution proportional to $(\chi^2/N)^4$ was added for these
experiments as discussed in Sec.\ \ref{subsec:ModifiedWeights}.  The dotted 
curve in 
Fig.\ \ref{fig:figEleven} shows the effect of setting all of the weight factors 
to 1 (including those for the jet experiments) and dropping the quartic penalty.
The resulting change is very small.  (The real purpose of the weights is toward 
maintaining acceptable fits to all experiments as we move away from the 
best fit to estimate uncertainties.)  

The short-dash curve in Fig.\ \ref{fig:figEleven}
shows the effect of dropping the Run I data, keeping the 
weights at 1, but restoring the quartic penalties on Run II jet $\chi^2$ values.
Finally, the long-dash curve is similar except that the quartic penalties have 
also been dropped.  This fit has weight 1 for all experiments except for dropping 
the older jet data, and no extra penalties added to $\chi^2$.  Some would argue 
this to be the most natural choice; though our belief is that it is preferable to 
apply some emphasis in the global fit to 
experiments that measure an important feature with a relatively small number 
of data points.  In any case, the uncertainty band shown is seen to do a 
reasonable job of encompassing the results of various plausible choices. 
If it were made much narrower by a smaller $\Delta\chi^2$ criterion, it would 
not do so.  \emph{Thus we see that a large part of the uncertainty---and the 
need for the $\Delta\chi^2 \sim 100$ criterion---arises from differences in 
plausible choices involved in making the global fit, rather than directly from 
propagating the experimental errors given in the data.}

\begin{figure}[tbh]
\begin{center}
 \resizebox*{0.40\textwidth}{!}{
 \includegraphics{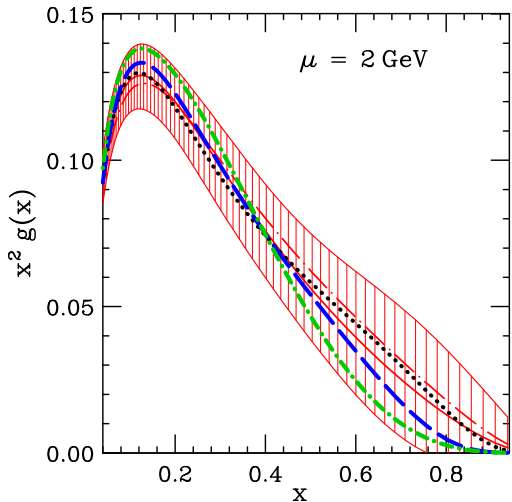}
 }
\end{center}
 \caption{
Solid curve and shaded region is CT09 ($\alpha_s(m_Z) = 0.118$).
All other curves are fits with $\alpha_s(m_Z) = 0.12018\,$. 
Dotted short dashed curve is MSTW2008NLO.
See text for description of the other curves. 
}
 \label{fig:figTwelve}
\end{figure}

\section{Comparison with MSTW}
\label{sec:CompareMSTW}
We compare our work with recent results from MSTW \cite{MSTW2008}
in Fig.\ \ref{fig:figTwelve}.  The solid curve and shaded region 
show the central fit and uncertainty range for CT09, as in the 
preceding figures.   
To make a straightforward comparison, all other curves in 
Fig.\ \ref{fig:figTwelve} use the MSTW values
$\alpha_s(m_Z) = 0.12018$, 
$m_c = 1.4 \, \mathrm{GeV}$, $m_b = 4.75 \, \mathrm{GeV}$.
The dotted long dashed curve is a fit that is the same as CT09 
except for the change in $\alpha_s(m_Z)$ (and the change in 
quark masses, which has a negligible effect).

The dotted short dashed curve is MSTW2008NLO.  It is surprisingly 
different from the $\alpha_s$-modified CT09, though it lies 
within our estimated 90\% confidence region.

To look for the cause of the difference between the $\alpha_s$-modified 
CT09 result and MSTW, we explore a series of modifications to the CT09
procedure that make it more like that of MSTW.  These are the 
same modifications that were discussed in connection with 
Fig.\ \ref{fig:figEleven}.  
First we drop the 
CDF$_{\mathrm{I}}$  and D0$_{\mathrm{I}}$ data sets.  This leads 
to the dotted curve in Fig.\ \ref{fig:figTwelve}, which is 
closer to the MSTW result at large $x$, but still quite far from it.

The dashed curve in Fig.\ \ref{fig:figTwelve} corresponds to 
again dropping the Run I data sets, while also setting the 
weight factors for all experiments to 1 and dropping the 
quartic penalties on $\chi^2/N$.  This reduces the influence 
of the jet data, and hence results in a fit that is closer 
to no-jets fits, which have a lower gluon at large $x$.
This dashed-curve fit is the most similar in its approach and result
to that of MSTW; but a noticeable difference 
still remains.  We can only speculate on what might 
be responsible for this difference, with obvious suspects being the 
different parametrizations used, or the neglect of correlated 
systematic errors for DIS data in the MSTW fit. 
Other possible sources for the difference is that there are some 
differences in which data sets are included in the fits, and a 
difference in the kinematical cuts in $Q$ and $W$ that are applied 
to those data sets.  Furthermore, there are small differences in the 
treatment of heavy quarks; and a small difference in the definition 
of $\alpha_s(\mu)$ at NLO, even when the values are matched at 
$\mu = m_Z$ (see \cite{cteqAB,stability}).

\begin{figure*}[tbh]
\begin{center}
 \resizebox*{0.85\textwidth}{!}{
 \includegraphics{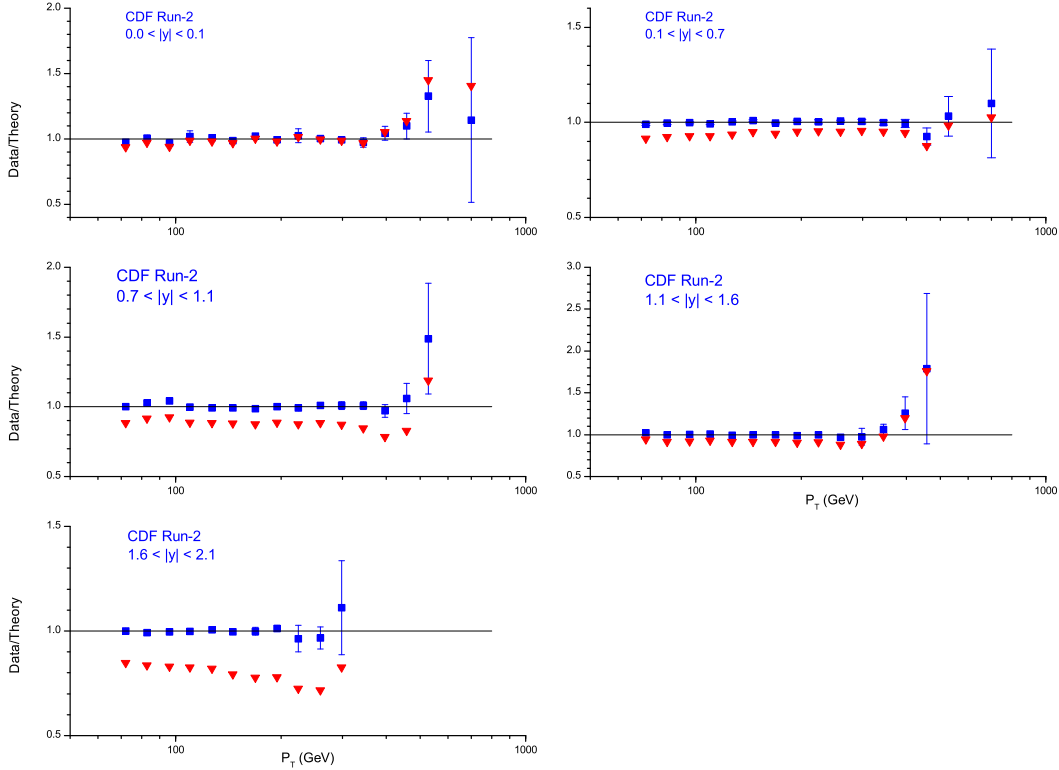}
 }
\end{center}
 \caption{Central fit to CDF Run II data. Triangles are the 
original data points. Squares with error bars include 
shifts due to the systematic errors, whose magnitudes are 
determined in the fit.}
 \label{fig:figThirteen}
\end{figure*}

\begin{figure*}[tbh]
\begin{center}
 \resizebox*{0.85\textwidth}{!}{
 \includegraphics{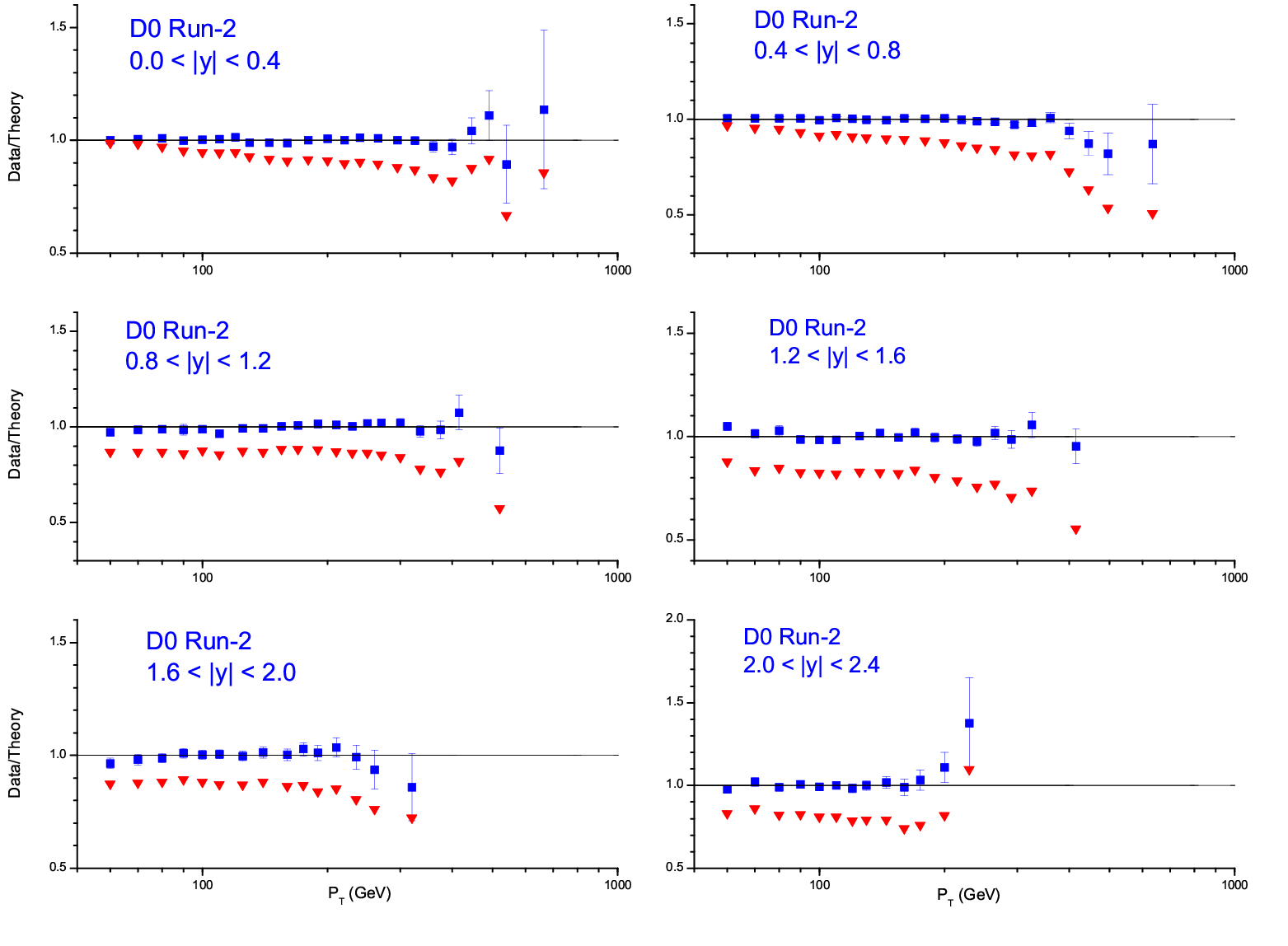}
 }
\end{center}
 \caption{Central fit to D0 Run II data. Triangles are the 
original data points. Squares with error bars include 
shifts due to the systematic errors, whose magnitudes are 
determined in the fit.}
 \label{fig:figFourteen}
\end{figure*}

\section{Quality of the fits}
\label{sec:FitQuality}

The good agreement of the central fits with the Run II jet data, when 
systematic error shifts allowed by the published data are included,
is shown in Figs.\ \ref{fig:figThirteen} and \ref{fig:figFourteen}. 
The unshifted data points are also shown.  These are quite far from 
the theory curves:  the systematic errors are much larger than the 
statistical ones here, so fitting the systematic error parameters is 
an essential part of fitting these data sets.  

There are 24 systematic shifts for CDF$_{\mathrm{II}}$, whose fitted values 
come out of order 1 as they should:
-0.1, -1.0, -0.3, -1.0, 0.7, -0.2, 0.8, -0.7, -0.7, -0.9, 0.1, 0.6,
 1.0, -0.3, -0.3, 0.5, -1.2, 0.4, 0.9, 0.0, -1.3, 0.1, -0.1, -0.3.
The fitted overall normalization factor is $1.02$, which is well within
the published 6\% error. 

There are 22 systematic shifts for D0$_{\mathrm{II}}$ (in addition to the 
overall normalization).  Some of these come out a bit larger, though they are 
still of order 1:
-0.5, -1.6, 0.0, 0.1, -0.8, -0.5, 0.1, 1.1, -0.4, 1.1, -1.1,
-0.4, 0.4, -1.6, -0.2, -1.9, 0.5, 0.3, 0.2, 1.7, -1.1, -0.1.
We presume these shifts to be reasonable, since their overall $\chi^2$ 
probability is acceptable, and since---as is typical of systematic 
errors---their experimental assessment must be partly subjective.  For what it 
is worth, we find it absolutely necessary for some of these shift parameters 
(most notably, ``dsys015: eta-intercalibration fit'') to have magnitude larger 
than 1.5 in order to achieve an acceptable fit to these data within the global 
fit.
The fitted overall normalization factor is $0.98$, which is 
well within the published error estimate. 

\section{Conclusion}
\label{sec:Conclusion}

We have carefully examined the NLO treatment of inclusive jet data and its 
influence on the determination of the gluon distribution in a QCD global 
analysis.  Key features of the analysis are 
the use of sufficiently flexible functional forms to reduce parametrization 
dependence, and full inclusion of the correlated systematic errors
published by the experiments.  

The difference between the new CT09 gluon results and our previous CTEQ6.6 
analysis \cite{cteq66} is shown in Fig.\ \ref{fig:figEight}.  
At a large scale like $\mu = 100 \, \mathrm{GeV}$, where the 
high-x gluon PDF is important for many high-profile signal and 
background processes at the Tevatron and LHC, the impact of the new 
jet data is quite small compared to the remaining uncertainty---as was 
expected from the outset, since the new data agreed fairly well with their  
prediction from CTEQ6.6 and its uncertainty range.

At the small scale $\mu = 2 \, \mathrm{GeV}$, where the constraints on 
the gluon are rather indirect, Fig.\ \ref{fig:figEight} shows that the 
change in the central prediction at some values of $x$ is close to the 
90\% confidence limit of the uncertainty estimated in CTEQ6.6.  
This demonstrates that our method does not 
overestimate those uncertainties, is spite of its tolerance for a range 
of $\chi^2$ that is large by ideal statistical standards.

We have introduced an extension of the familiar Hessian matrix 
method \cite{Hessian} for uncertainty analysis. The extension involves 
making a further orthogonal transformation of the coordinates, after 
the transformation that diagonalizes the Hessian has been carried 
out.  This leaves the Hessian matrix in its convenient diagonal form, 
while offering the possibility to describe the uncertainty on a given 
quantity using a small number of important eigenvector sets.  This is 
illustrated in the right-hand side of Fig.\ \ref{fig:figSix}, where most 
of the gluon uncertainty near $x=0.5$ is given by just one or two 
eigenvector pairs.  
A further application of this extension of the Hessian method 
provides a new and improved method to study the compatibility of the
data sets in a global fit.  This is described in a separate 
publication \cite{Measuring}.

One value of this paper is to document and illustrate methods that can be 
used to incorporate new data sets into a global analysis.  There will be many 
opportunities to apply this in the near future, as data from Tevatron Run II and 
HERA Run II continue to arrive, and with data from the LHC on the horizon.

\begin{figure*}[tbh]
\begin{center}
 \resizebox*{\textwidth}{!}{
\mbox{
\resizebox{0.25\textwidth}{!}{
\includegraphics[clip=true,scale=1.0]{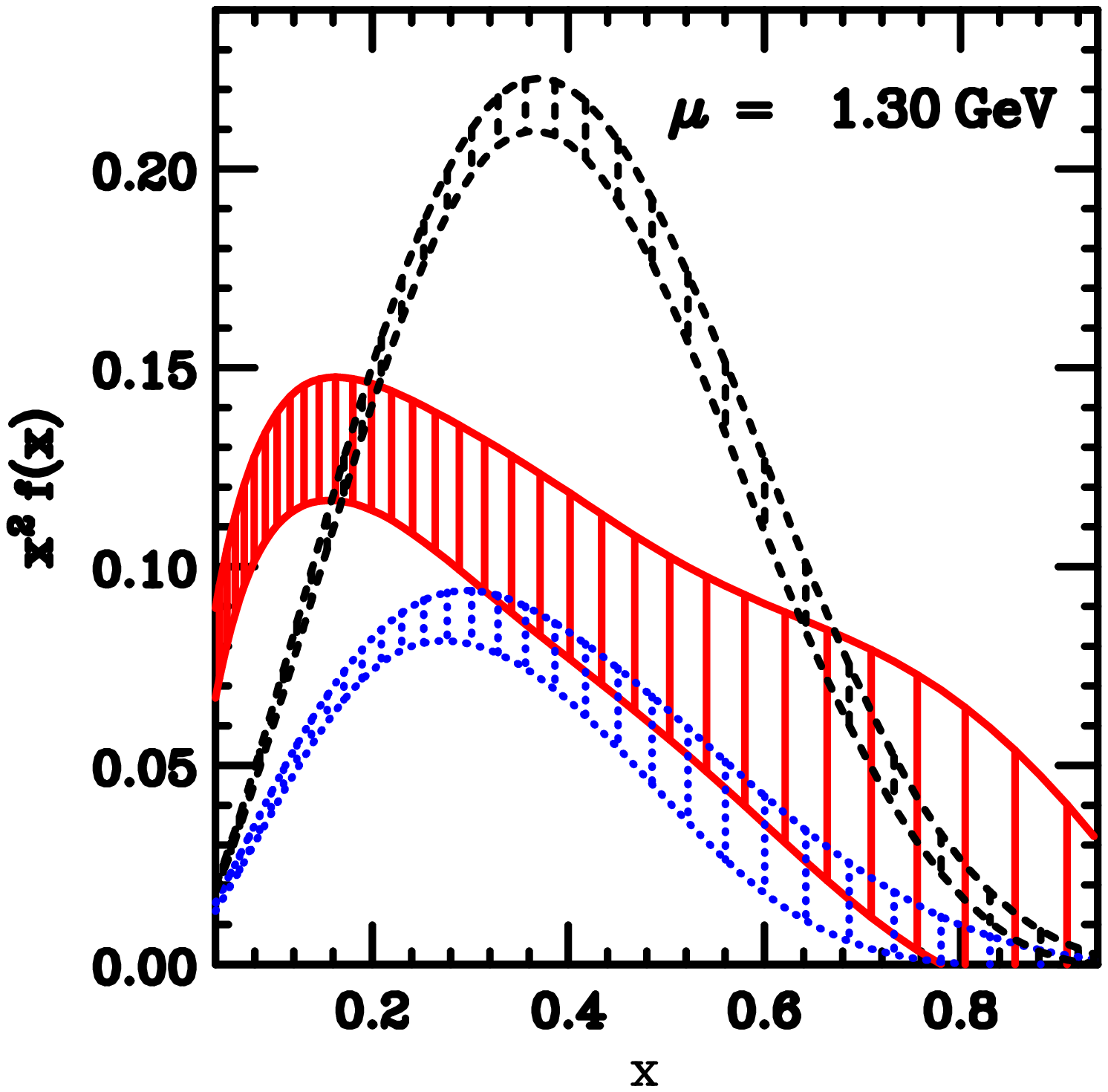}}
\hfill
\resizebox{0.25\textwidth}{!}{
\includegraphics[clip=true,scale=1.0]{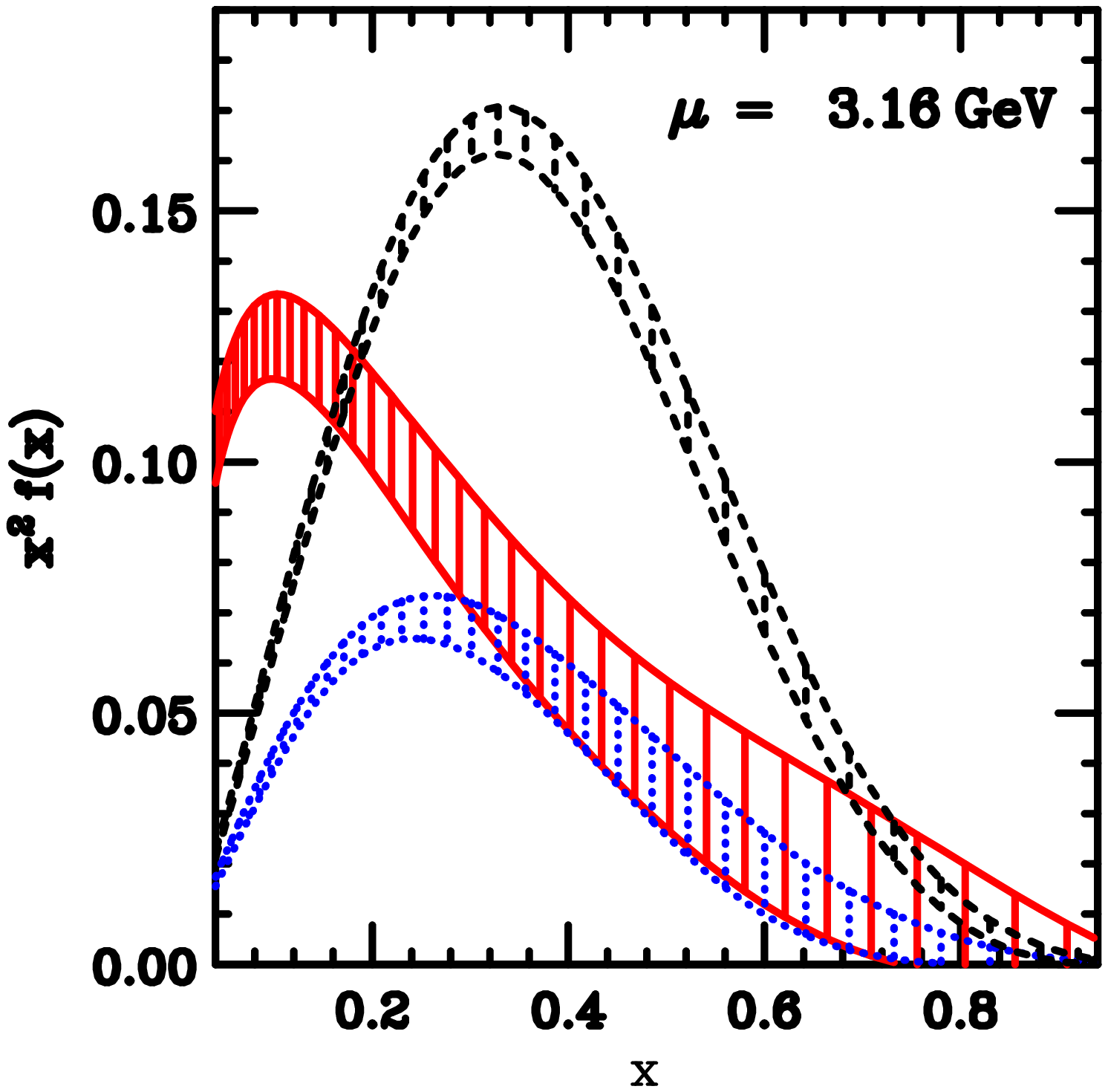}}
\hfill
\resizebox{0.25\textwidth}{!}{
\includegraphics[clip=true,scale=1.0]{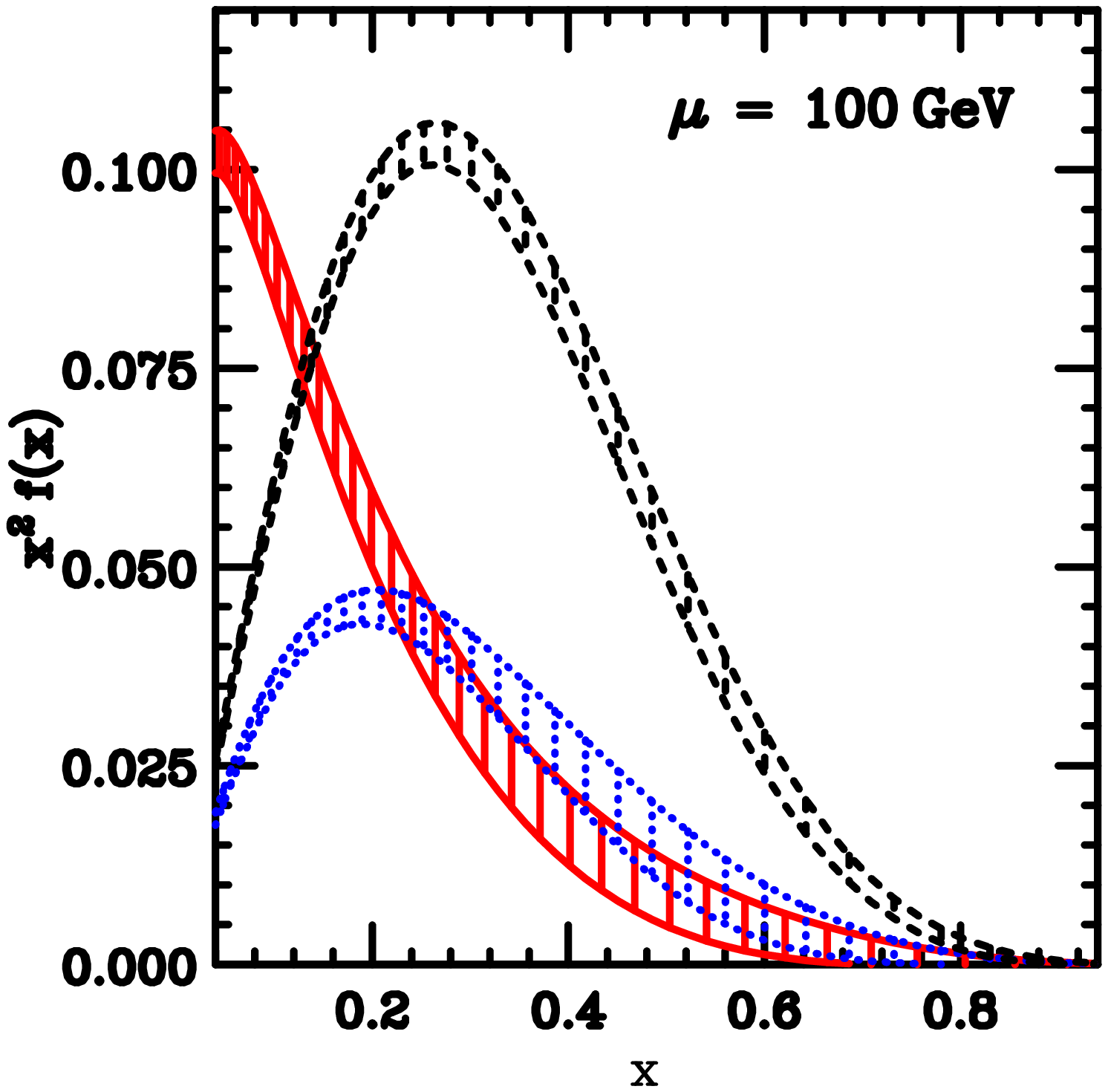}}
}
 }
\end{center}
 \caption{Gluon (solid),  $u$ quark (dashed),
and $d$ quark (dotted) distributions at three different scales.
}
 \label{fig:figFifteen}
\end{figure*}

To conclude with a speculation, it is interesting to compare the extracted 
gluon distribution with the distributions for 
up and down quarks.  This comparison is shown in 
Fig.\ \ref{fig:figFifteen}.  The quark distributions have smaller uncertainties than 
the gluon---particularly the up quark, whose larger electric charge makes it
prominent in the extensive body of neutral-current DIS measurements.
Surprising as it may seem, we observe that at a small scale like 
$\mu = 1.3 \, \mathrm{GeV}$, the gluon 
PDF is most likely larger than the down quark distribution even at very large $x$.
It may or may not even be larger than the up quark distribution---more data will
be needed to determine that.  
An important challenge for further study would be to see if perhaps one can argue
convincingly from models of the nonperturbative physics of the proton that 
``valence-like'' gluon alternatives where $g(x,\mu_0) > u(x,\mu_0)$ at large $x$ 
are unphysical, in which case the uncertainty in PDFs could be significantly 
reduced.

\bigskip
\emph{In Memoriam:}
It has been our pleasure to work with, and be inspired by, our late mentor,
colleague, and friend Wu-Ki Tung. Much of the methodology of modern PDF global 
analysis was his innovation, and he remained involved in this work to the end 
of his life.

\begin{acknowledgments}
We thank D.~Soper for discussions on the EKS NLO jet program.
We thank Kenichi Hatakeyama for providing the CDF Run II data files.
We thank M.~Ubiali, A.~Guffanti, S.~Forte, and J.~Rojo for discussions 
on the NNPDF approach.  The revised version of the paper benefited from 
insightful suggestions from the anonymous referee for Phys.\ Rev.\ D.

This research was supported by the 
U.S. National Science Foundation under grants PHY-0354838, PHY-055545, 
and PHY-0757758; 
U.S.\ Department of Energy under grant
DE-FG02-04ER41299; 
National Center for Theoretical Sciences and National Science Council
of Taiwan under grant NSC-97-2112-M-133-001; and by the Lightner-Sams Foundation.

\end{acknowledgments}

\section*{Appendix: Alternative choices for eigenvectors \label{sec:rediag}}

Here we sketch how the eigenvector PDF sets in a global
QCD analysis can be recalculated to more simply represent the
uncertainty of a particular physics quantity, such as the
gluon distribution at large $x$ that is studied in this paper. 
See \cite{Measuring} for a more detailed description.

The standard Hessian method for error analysis is based on a quadratic expansion 
of $\chi^{2}$ in the neighborhood of the minimum of a global fit.  This expansion 
follows from Taylor series: 
\begin{equation}
\chi^{2} \, = \, \chi_{0}^{2} \, + \, 
 \frac{1}{2} \sum_{i=1}^N \sum_{j=1}^N 
 \left(\frac{\partial^{\, 2}\chi}{\partial a_i\,\partial a_j}\right)_0  
 (a_i-a_j^{(0)}) \, (a_j-a_j^{(0)}) \;,
\label{eq:chi1}
\end{equation}
where there are no first-order terms because the expansion is about the minimum, 
and terms higher than second order have been dropped.  The $\{a_i\}$ in 
Eq.\ (\ref{eq:chi1}) are the parton parameters of the global fit, and quantities 
with superscript $(0)$ are evaluated at the minimum of $\chi^2$.
Formally, one can express the displacements
$a_i-a_i^{(0)}$ as linear combinations of the normalized eigenvectors of 
the matrix of second derivatives to obtain a diagonal expression
\begin{equation}
\chi^{2} \, = \, \chi_{0}^{\,2} \, + \, \sum_{i=1}^N z_{i}^{\,2}
\label{eq:chi2}
\end{equation}
in which the new coordinates $\{z_i\}$ are the coefficients that multiply the 
eigenvectors.
Because nonquadratic behavior appears at widely different scales in 
different directions of the parameter space, and because the 
second-derivative matrix must be calculated numerically by finite differences, 
it is necessary in practice to compute the linear transformation from 
coordinates $\{a_i-a_i^{(0)}\}$ to coordinates $\{z_i\}$ by a series 
of iterative steps \cite{multivariate}.

The choice of eigenvectors that define the transformation to the diagonal 
form (\ref{eq:chi2}) is not unique, because any further orthogonal 
transformation of the parameters $\{z_{i}\}$ will preserve that form.  
This freedom to make a further orthogonal transformation can be used to 
simultaneously diagonalize any one additional function of the coordinates 
within the quadratic approximation.  Specifically, if $G$ is a function 
of the original coordinates, one can choose the new coordinates such that
\begin{equation}
G \, = \, G_0  \, + \, \sum_{i=1}^N (P_i \, z_i \, + \, Q_i \, z_i^{\, 2}) \; .
\label{eq:Gappendix}
\end{equation}
while maintaining (\ref{eq:chi2}).  This form (\ref{eq:Gappendix}), which is 
accurate through second order in the $\{z_{i}\}$, is obtained by the 
following recipe:  (1) Calculate the symmetric matrix 
$(\partial^{\,2}G/\partial z_i\,\partial z_j)_0$ using the ``old'' 
$\{z_{i}\}$ by finite differences; (2) Express these ``old'' $\{z_{i}\}$ 
as linear combinations of the eigenvectors of that matrix; (3) The coefficients 
of these linear combinations become the desired ``new'' $\{z_{i}\}$.  
These steps are iterated a few times to refine the transformation.
This procedure is described explicitly in \cite{Measuring}.

In the iterative procedure used in our previous uncertainty 
analyses \cite{cteq606165, cteq66, Hessian}, the quantity $G$ defining 
the transformation (\ref{eq:Gappendix}) was the overall length-squared 
of the 
displacement from the minimum in the space of the original shape 
parameters: $\sum_{i=1}^N (a_i-a_i^{(0)})^{2}$.
To study the uncertainties of
$g(x,\mu)$ at large $x$, we can instead choose $G$ to be a Mellin
moment of some PDF, such as 
$\int_{0}^{1}x^{n}g(x,\mu_0)\,dx$ with $2 \lesssim n \lesssim 5$; or 
we can simply choose $G = g(x,\mu_0)$, e.g., at $x = 0.55$ as was done to
create the right hand side of Fig.\ \ref{fig:figSix}.  To facilitate the 
study of some interesting physical quantities, one might want to choose $G$ 
to be, say, the cross section for $W$, $Z$, or Higgs boson production.  
Another choice, which is useful for exploring the internal consistency of 
a global fit, is to define $G$ as the contribution to $\chi^2$ 
from a particular subset of the data.  This application is the subject 
of \cite{Measuring}.

\end{document}